\documentclass[twocolumn,aps,floatfix,showpacs]{revtex4}

\usepackage{amsmath}
\usepackage{amssymb} 
\usepackage{graphics}

\begin{document}

\newcommand\UBA{{\sc uba}}
\newcommand\RBN{{\sc rbn}}
\newcommand\EP{{\sc ep}}
\newcommand\SP{{\sc sp}}
\newcommand\trm\textrm
\newcommand\mbf\mathbf
\newcommand\s{\mbf{s}}
\newcommand\kb{\mbf{k}}
\newcommand\Nt{\tilde N}
\newcommand\pt{\tilde p}
\newcommand\ut{\tilde u}
\newcommand\yt{\tilde y}
\newcommand\tti{\tilde t}
\newcommand\Nh{\hat N}
\newcommand\uh{\hat u}
\newcommand\rh{\hat r}
\newcommand\ti{\trm i}
\newcommand\tn{\trm n}
\newcommand\tu{\trm u}
\newcommand\tsa{{0*}}
\newcommand\tsia{{0,0*}}
\newcommand\msa{\trm\tsa}
\newcommand\msia{\trm\tsia}
\newcommand\Pex{P_{\trm{\sc cc}}}
\newcommand\PNex{P_{\trm{\sc cc}}^1}
\newcommand\Pn{P_{\tn,N}}
\newcommand\Pu{P_{\tu,N}}
\newcommand\Oc{{\cal O}}

\title{Exhaustive percolation on random networks}

\author{Bj\"orn~Samuelsson} 
\email[]{bjorn.samuelsson@duke.edu}
\author{Joshua~E.~S.~Socolar} 
\email[]{socolar@phy.duke.edu}
\affiliation{Physics Department and Center for Nonlinear and Complex
  Systems, Duke University, Durham, NC, 27514}

\date{\today}

\begin{abstract}
  We consider propagation models that describe the spreading of an
  attribute, called ``damage'', through the nodes of a random network.
  In some systems, the average fraction of nodes that remain undamaged
  vanishes in the large system limit, a phenomenon we refer to as {\it
    exhaustive percolation}.  We derive scaling law exponents and
  exact results for the distribution of the number of undamaged nodes,
  valid for a broad class of random networks at the exhaustive
  percolation transition and in the exhaustive percolation regime.
  This class includes processes that determine the set of frozen nodes
  in random Boolean networks with indegree distributions that decay
  sufficiently rapidly with the number of inputs.  Connections between
  our calculational methods and previous studies of percolation
  beginning from a single initial node are also pointed out.  Central
  to our approach is the observation that key aspects of damage
  spreading on a random network are fully characterized by a single
  function specifying the probability that a given node will be
  damaged as a function of the fraction of damaged nodes.  In addition
  to our analytical investigations of random networks, we present a
  numerical example of exhaustive percolation on a directed lattice.
\end{abstract}


\maketitle

\section{Introduction}
\subsection{Overview}
Propagation models on lattices or more general graphs describe the
spreading of some discrete signal through a set of discrete
entities.  In the most general terms, the signal corresponds to some
qualitative change that causes the entity to interact differently with
its neighbors.  Examples include the spreading of damage in power
grids \cite{Sachtjen:00, Kinney:05}, the spreading of disease through
a population \cite{Hethcote:00, Newman:02, Cohen:03}, the spreading of
a computer virus on the Internet \cite{Pastor-Satorras:01, Lloyd:01},
or the alteration of gene expression patterns in a cell due to a
mutation \cite{Hughes:00, Ramo:06}.  In the general case, the
individual entities are represented as nodes in a graph where the
links indicate paths along which the signal can spread
\cite{Watts:99,Newman:03, Callaway:00, Moreno:02, Watts:02, Motter:04,
Crucitti:04}.  Because the signal can be thought of as disrupting the
static or dynamical state of the original system, we refer to its
propagation as spreading {\it damage}, though in many cases the ``damage''
may enhance a desired property or simply represent some natural
dynamical process.  A single instance of a given spreading process
initiated from a particular subset of nodes is often called an
avalanche.

In analyzing spreading processes, one is often interested in the
transition between those that die out quickly and those that spread to
a finite fraction of the system in the large-system limit, a
transition that may occur as the probability of transmitting damage
across links is varied.  This percolation transition is relevant for
systems in which the fraction of initially damaged nodes tends to
zero in the limit of infinite system size.  The order parameter for
the transition is the average fraction of nodes damaged in a single
avalanche, which remains zero for small transmission probabilities and
continuously increases when the probability rises above a threshold
value.  We will refer to this as the {\it sparse percolation} (\SP)
transition.  The \SP\ transition occurs for spreading processes in
which the probability that a node becomes damaged is zero unless at
least one of its neighbors is damaged.  (If this probability were
nonzero, a nonzero fraction of the nodes would always get damaged.)

For a certain class of propagation models, there is another transition
of interest.  When the fraction of initially affected nodes remains
fixed as the system size is increased, the fraction of nodes that
remain {\it undamaged} can undergo a transition from finite values to
zero at transmission probabilities above some threshold.  We refer to
this as the {\it exhaustive percolation} (\EP) transition.  The \EP\ 
transition occurs only for propagation models in which the probability
of a node remaining undamaged is zero when all of its neighbors are
damaged (all of its inputs in the case of a directed graph).  We
assume also that there is a nonzero probability for a node to remain
undamaged if it has at least one undamaged input.  There is then one
more condition for the \EP\ transition: the density of directed loops
of any specified size must vanish in the large system limit.  For any
loop there is a finite probability that no member of the loop will be
damaged, since no member of the loop can have all of its inputs
damaged until one of the members becomes damaged through a
probabilistic event.  Thus \EP\ is {\it not} observable on spatial
lattices of the type generally encountered in statistical mechanics.
\EP\ is observable, however, on directed lattices and on graphs in
which the nodes serving as inputs to a given node are selected at
random.

In this paper we derive the probability distribution for the number of
undamaged nodes at the \EP\ transition on random graphs for a general
class of propagation models exhibiting what we call {\it unordered
  binary avalanches} (\UBA).  This is analogous to finding the
distribution of avalanche sizes at the usual percolation transition,
but here we are asking for the distribution of the number of nodes
{\it not} participating in the avalanche.

As an application of our \EP\ results, we consider the problem of
identifying unfrozen nodes in a random Boolean network (\RBN).  In a
\RBN, each node has a binary state that is updated according to a rule
that takes the values of some other nodes as inputs. The dynamics of
\RBN s has been investigated extensively; see, e.g.,
\cite{Kauffman:69, Derrida:86a, Bastolla:97, Socolar:03, Aldana:03a,
  Aldana:03b, Samuelsson:03, Kauffman:04, Drossel:05b}.  A \RBN\ can
have several dynamical attractors, but some nodes might have the same
value at all times on all attractors. Such nodes are called {\it
  stable} and the set of stable nodes is important for the dynamics in
\RBN s \cite{Flyvbjerg:88b, Bastolla:98a, Bilke:01}.

Almost all stable nodes in a broad class of \RBN s can be identified
through a dynamic process that was introduced by Flyvbjerg
\cite{Flyvbjerg:88b} and formalized to facilitate numeric simulations
by Bilke and Sjunnesson \cite{Bilke:01}.  We call the stable nodes
that can be identified by this dynamic process {\it frozen} (and nodes
that are not frozen are called {\it unfrozen}).  Provided that the
Boolean rule distribution is symmetric with respect to inversion of
any subset of inputs, the set of frozen nodes can be identified
through an \UBA\ in which frozen inputs cause new nodes to become
frozen (damaged).  Most rule distributions that have been examined in
the literature exhibit this symmetry.  The requirement is satisfied,
for example, for any model that assigns given probability $p$ for
obtaining a 1 in each entry of the truth table for each node.  

This paper is organized as follows.  We first develop the notation and
basic definitions required for discussing \UBA s in general.  In
Section~\ref{sec: basic defs}, we give an introduction to the \UBA\ 
formalism from the perspective of percolation processes.  A more
formal description is given in Section~\ref{sec: introduction to EP},
followed by a numerical illustration of the basic concepts.  In
Section~\ref{sec: random networks}, we present analytic derivations
for \UBA\ in random networks with emphasis on \EP\ and the \EP\ 
transition.  We also present explicit results for the special case of
Erd\H{o}s--R\'{e}nyi networks with a natural choice for the avalanche
rules.

In Section~\ref{sec: application} we show how to apply the \UBA\ 
formalism to obtain the statistics of frozen nodes in two-input \RBN
s.  In the present context, this serves as an illustration of the
general theory, but this particular example was also the primary
motivation for studying \EP.  The results on \RBN s are consistent
with those found by Kaufman, Mihaljev, and Drossel.~\cite{Kaufman:05}.
The main advantage of using the \EP\ formalism for this problem is
that it makes clear how the calculation can be extended to networks
with more than two inputs per node, including networks with an
in-degree distribution that (with a low probability) allows
arbitrarily large in-degrees.

\subsection{Basic definitions}
\label{sec: basic defs}

An unordered binary avalanche (\UBA ) is defined as a spreading process
with the following properties:

\begin{description} 
  
\item[Binary states:] the state of each node can be characterized as a
  binary variable $s$, with $s=0$ meaning {\it undamaged} and $s=1$
  meaning {\it damaged};
  
\item[Boolean rules:] the state of each node is determined by a
  Boolean function of the states of its input nodes;
  
\item[Order independence:] the probability of having a given set of
  nodes damaged at the end of the process does not depend upon the
  order in which nodes are chosen for updating.

\end{description}

Order independence refers to the dynamics of the spreading process or
a simulation of it.  In such a simulation, one typically chooses a
site and updates it according to a rule depending on the states of
sites that provide inputs to it, repeating the process until a test of
every site yields no change in the state of the system.  We are
interested in cases where the order in which sites are chosen for
possible updating has no bearing on the final state of the system.

\UBA\ is a natural extension of site or bond percolation.  To determine
the avalanche size distribution in site percolation, for example, one
identifies an initial subset of damaged sites and then tests neighbors
of damaged sites to see whether the damage spreads to them.  After a
given site is tested for the first time, its value is permanently
fixed.  The process is iterated until no new damaged sites are
generated.  See, e.g., Ref.~\cite{Sahimi:03}.  This method of
investigating site percolation is equivalent to assigning all sites a
value, then beginning with a damaged site and determining all of the
damaged sites in a connected cluster.  Site percolation where each
site has the probability $p$ to be occupied can be recast as a \UBA\
system as follows.  Let each site be associated with a rule that is an
{\sc or}-rule of all of its neighbors with probability $p$ and is a constant
0 with probability $1-p$. Then the above described site percolation
is achieved by first selecting the rules and clamping the value of a
given site to 1, and then repeatedly updating the system according to
the Boolean rules.  In this situation, the 1s in the final state mark
a site percolation cluster.  A more practical way of simulating the
same \UBA\ is to determine probabilistically the Boolean rule at each
site only when that site is first encountered in the percolation
process and to update only those nodes where the rules have been
determined.

To ensure order independence in \UBA, it is sufficient to require that
each Boolean function is non-decreasing, meaning that if one of the
inputs to the rule changes from 0 to 1, the output is not allowed to
change from 1 to 0.  For non-decreasing Boolean functions, if a
specific node is eventually going to be assigned the value 1 during an
avalanche, updating other nodes to 1 first cannot change the outcome.

We are particularly interested in \UBA s that are initiated by damage
at a set of nodes comprising a nonzero fraction of the total number of
nodes.  Such a process would be relevant, for example, if the
probability that any given node is damaged at the start is independent
of the system size.

To clarify both the distinction between \EP\ (exhaustive percolation)
and \SP\ (sparse percolation) and the similarities between them, we
describe a particular case of a propagation model that exhibits both
transitions.  Consider a graph with a total of $N$ nodes, some of
which have three input links each while the others have no input links
at all.  The graph is random in that the node supplying the input
value on any given link is selected at random, but stays fixed
throughout the avalanche.  Let $\nu_0$ be the fraction of nodes with
no inputs.  Define a spreading process as follows: The initial
condition is that all nodes with no inputs are considered damaged.
Each other node is now selected in turn to see whether the damage
spreads to it.  If a node has one damaged input, the probability that
it will be damaged is $p_1$; if it has two damaged inputs, the
probability of damage is $p_2$ (with $p_2 \ge p_1$); and nodes with
three damaged inputs are guaranteed to become damaged ($p_3 = 1$).
These probabilities are realized, for example, by the following
Boolean rule distribution: a 3-input {\sc or}-rule with probability
$p_1$; a 3-input majority rule with probability $p_2-p_1$; and a
3-input {\sc and}-rule with probability $1-p_2$.

As $N$ goes to infinity, the number of initially damaged nodes can be
a nonzero number that grows slower than $N$, meaning that
$\nu_0$ goes to zero as $N$ goes to infinity.  In this limit, the
\SP\ transition occurs at $p_1 = 1/3$ and the spreading from each
initially damaged node is described by a {\it Galton--Watson process}.
In a Galton--Watson process, a tree is created by adding branches to
existing nodes, with the number of branches emerging from each node
drawn from a fixed probability distribution.  Such branching processes
have been investigated extensively. (See, e.g.,
Ref.~\cite{Harris:63}.)  In particular, the correspondence to
Galton--Watson processes means that for critical \SP, the probability
of finding $n$ damaged nodes scales like $n^{-3/2}$ for $1\ll n\ll N$
\cite{Otter:49, Ramo:06}.

For any nonzero value of $\nu_0$, the \EP\ transition occurs for $p_2$
satisfying $(1 - p_2)(1-\nu_0) = 1/3$ (assuming that this value of
$p_2$ is greater than $p_1$.)  The analysis described in
Section~\ref{sec: random networks} provides a
method of calculating the probability $P(u)$ of having $u$ {\it
  undamaged} nodes in this case.  The result in the large $N$ limit is
$P(u) \sim P(0)u^{-1/2}$ for large $u$.  A difference between \EP\ and
\SP\ is that both $P(0)$ and the cutoff on the $u^{-1/2}$ distribution
scale with $N$ for \EP, while for \SP\ only the cutoff scales with
$N$.

\section{Introduction to exhaustive percolation}
\label{sec: introduction to EP}

\subsection{Formal description of UBA}
\label{sec: formal UBA}

We now describe a formalism and establish some notation that is
suitable for a detailed treatment of \UBA.  Let $N$ denote the number
of nodes in a network with a specified set of links and let the nodes
be indexed by $j=1,\ldots,N$.  The network state is described by the
vector $\s=\{s_1,\ldots,s_N\}$.  Let $K_j$ denote the number of inputs
to node $j$, and let $\kb_j$ denote the vector of $K_j$ inputs to node
$j$.  Furthermore, let $R$ denote a Boolean function and let
$\Pi_j(R)$ denote the initial probability that node $j$ has the rule
$R$. [It is required that $R$ has precisely $K_j$ inputs for
$\Pi_j(R)$ to be nonzero.]

To efficiently simulate \UBA, we keep track of the information that is
known about each node at each step in the process.  In
particular, it is important to keep track of whether or not the change
from 0 to 1 of a given input has already been accounted for in
determining the output.  The simplest way to do this is to introduce
an extra state \tsa\ that labels a site whose rule $R$ implies an
output value of 1 but for which the update to 1 has not yet been
implemented.  When a node changes its state from 0 to \tsa, it is a
silent change in the sense that the Boolean rules at the other nodes
treat an input \tsa\ exactly the same as 0. To retrieve the final
state of the network, all occurrences of \tsa\ must be updated to 1.
When a single update to 1 is made, the information that the given node
has value 1 is passed along to all nodes with inputs from it.  The
values of these nodes may then change from 0 to \tsa.  The conditional
probability that the value of node $i$ is updated from 0 to \tsa\ when
$j$ changes value from \tsa\ to 1, is given by
\begin{equation}
  U_i(\s,j) \equiv \frac{P_1(\kb'_i) - P_1(\kb_i)}{1 - P_1(\kb_i)},
\label{eq: Ui}
\end{equation}
where $\kb'_i$ is the value of $\kb_i$ after $s_j$ has been updated and
$P_1(\kb_i)$ is the probability that $R_i(\kb_i) = 1$:
\begin{equation}
  P_1(\kb_i) \equiv  \sum_R R(\kb_i)\Pi_i(R).
\label{eq: P1}
\end{equation}
The numerator in Eq.~(\ref{eq: Ui}) is the probability that $R_i$
produces a 1 after the update of node $j$ minus the probability that
$R_i$ produced a 1 before the update.  The denominator is the
probability that node $i$ had the value 0 before the update.

Let $\Pi_i(1)$ denote the probability that the rule at node $i$ has
output 1 regardless of its input values. If some particular nodes are
selected for initiation of the \UBA, $\Pi_i(1)$ is set to one for
these nodes [which means $\Pi_i(R) = 0$ for all other rules].

We are now ready to present a formal algorithm for determining the
final state of an instance of \UBA\ on a finite network.  We carry out
the following procedure (where $:=$ denotes the assignment operator):
\begin{enumerate}
\item $s_j:=0$ for all $j$; 
\item $s_j:=\msa$ with probability $\Pi_j(1)$ for each $j$; 
\item Some $j$ with $s_j=\msa$ is selected; \label{j select}
\item $s_i:=\msa$ with probability $U_i(\s,j)$ for each $i$ with $s_i=0$; 
\item $s_j:=1$; \label{j := 1}
\item Steps \ref{j select}--\ref{j := 1} are iterated as long as there
exists a node in state \tsa. \label{iterate}
\end{enumerate}

\UBA\ can also be considered on infinite networks, but that requires a
more technical description of the process. First, the choices of $j$ in
step 3 for both descriptions must be such that any given $j$ that
satisfies the conditions in step~\ref{j select} will be selected in a
finite number
of iterations.  Second, the ensemble of final states needs to be
defined in terms of a suitable limit process because the stopping
criterion in step~\ref{iterate} can not be applied to an infinite
system.

Note that the dynamics is only dependent on the probability functions
$\{P_1(\kb_i)\}$.  That is, the precise rule distributions affect the
avalanche results only through their contributions to $P_1$.  Because
the Boolean rules are non-decreasing functions, $P_1(\kb_i)$ is also a
non-decreasing function. In fact, every non-decreasing function,
$f(\kb_i)$, with values in the interval $[0,1]$ can be realized by
$P_1(\kb_i)$ for a suitable Boolean rule distribution.  One such rule
distribution can be constructed as follow: for each $i$ and each
$\kb_i$, select a random number $y$ from a uniform distribution on the
unit interval and set $R_i(\kb_i) = 1$ if and only if $y < f(\kb_i)$.

\subsection{An example of EP on a lattice}
\label{sec: EP lattice}

To illustrate the concepts of \UBA\ and \EP, consider a directed
network on a two-dimensional square lattice with periodic boundary
conditions.  Each node in the lattice has integral coordinates $(i,j)$
where $i+j$ is odd and the node at $(i,j)$ receives inputs from the
two nodes at $(i-1,j\pm1)$.  The rule for propagation of damage to a
node is either {\sc or} or {\sc and}, with probabilities
$\Pi_{(i,j)}(\trm{\sc or}) = r$ and $\Pi_{(i,j)}(\trm{\sc and}) = 1 -
r$, respectively.

\begin{figure}[bt]
\begin{center}
\includegraphics{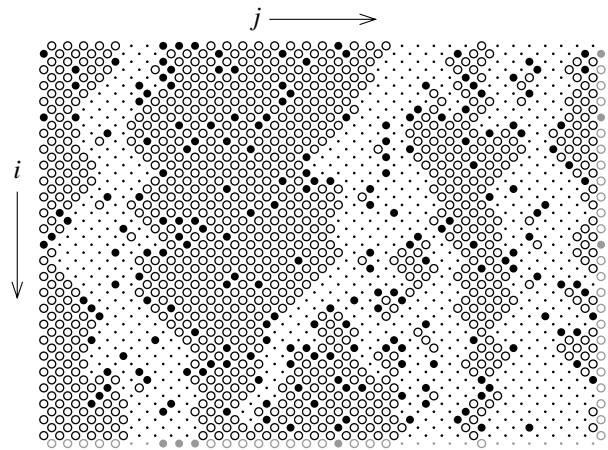}
\end{center}
\caption{\label{fig: lattice UBA} An example of \UBA\ on a lattice,
  displaying undamaged nodes (dots), initially damaged nodes (filled
  circles), and nodes damaged during the avalanche (empty circles).
  Each node has either an {\sc or}-rule or an {\sc and}-rule with
  inputs from its neighbors in the row immediately above the node.
  The probability for a node to be initially damaged is
  $\rho = 1/8$ and the probability for obtaining an {\sc
    or}-rule is $r=0.3$. Periodic boundary conditions are used and the
  first row and column are repeated in gray after the last row and
  column to illustrate the periodic boundary conditions. }
\end{figure}

Figure~\ref{fig: lattice UBA} displays an avalanche
that is initiated by letting each node be initially damaged with
probability $\rho=1/8$. A node assigned {\sc or} becomes
damaged if either of its neighbors one layer above is damaged; a node
assigned {\sc and} becomes damaged if and only if both neighbors above
it are damaged.  This means that
\begin{align}
  P_1(\kb_{(i,j)}) &= \left\{
  \begin{array}{ll}
     0 & \trm{if }\kb_{(i,j)} = (0,0)\\
     r & \trm{if }\kb_{(i,j)} \in \{(0,1),(1,0)\}\\
     1 & \trm{if }\kb_{(i,j)} = (1,1)~.
  \end{array}\right.
\end{align}
Note that clusters of damaged nodes formed in an avalanche initiated
by a single damaged node cannot contain any holes, as the uppermost
undamaged node in the hole would have to have two damaged inputs and
hence would become damaged when updated. 

For localized initial damage, the \SP\ threshold is found at $r =
1/2$.  Above this value of $r$, domains of damage tend to widen as the
avalanche proceeds.  Since the growing cluster has no holes, this is
simultaneously an \EP\ transition.  The \EP\ transition can be found
for smaller values of $r$ in lattices where each node is initially
damaged with a given nonzero probability $\rho$. [For every
initially damaged node, $\Pi_{(i,j)}(1)$ is set to 1, meaning that
$P_1(\kb_{(i,j)})=1$ for every value of $\kb_{(i,j)}$.]

\begin{figure}[bt]
\begin{center}
\includegraphics{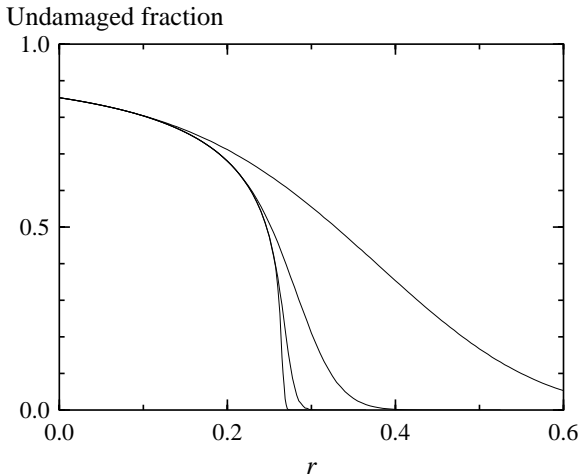}
\end{center}
\caption{\label{fig: lattice EP} The average fraction of undamaged
  nodes for \UBA\ on a lattice of the type shown in Fig.~\ref{fig:
    lattice UBA}, as a function of the selection probability $p$ for
  {\sc or}-rules and the probability $\rho=1/8$ for initial
  damage.  The lattice has periodic boundary conditions and covers a
  square that has a side of $10$, $10^2$, $10^3$, and $10^4$ lattice
  points, respectively, with steeper curves for larger systems. The
  statistical uncertainty in the estimated mean is less than the line
  width. }
\end{figure}

Figure~\ref{fig: lattice EP} shows the average number of unaffected
nodes as a function of $r$ for $\rho=1/8$ on lattices with
periodic boundary conditions.  The numerics displayed in
Fig.~\ref{fig: lattice EP} clearly suggest that there is a
second-order \EP\ phase transition.  Furthermore, these numerical
results suggest that the avalanche in Fig.~\ref{fig: lattice UBA} is
within the parameter regime for \EP\ and that \EP\ does not occur in
this case due to finite size effects.

For the case $r = 0$, it is possible to map the \EP\ transition onto
ordinary, directed, site percolation on the same lattice.  When all
nodes in the lattice have {\sc and}-rules, the following algorithm may
be used to determine whether a given node will be damaged: select a
node; put a mark on the selected node unless it is initially damaged;
and recursively mark each initially undamaged node that has an output
to a marked node.  The selected node will get damaged if and only if
this recursion ends in a finite number of steps.  The algorithm
describes ordinary directed site percolation where the initially
undamaged (damaged) nodes are considered active (inactive) sites and
the process propagates in the opposite direction relative to the \UBA.
We therefore expect the \EP\ transition to  occur for a value of
$\rho$ equal to $1-p_{\trm c}$, where $p_{\trm c} = 0.70549$ is the threshold for
directed site percolation \cite{Essam:88} and we have confirmed this
with numerical tests.  Further study of \EP\ on the lattice is beyond
the scope of this paper.

\subsection{Suppression of EP by resistant motifs}
\label{sec: resistant motifs}

In the lattice example above, the fact that the network had no
feedback loops smaller than the lattice size was important.  In
general, \EP\ is suppressed by the presence of short feedback loops.
As already noted, for \EP\ to occur, it is required that the output of
each rule in the rule distribution is 1 if all of its inputs have the
value 1.  Otherwise, there would be a finite fraction of nodes that
keep the value 0 regardless of the influence from the rest of the
network.  Generalization of this reasoning allows us to rule out \EP\
in other situations, indicating that \EP\ is most likely to occur in
directed or highly disordered networks. To pursue this idea, we
introduce the notion of {\it resistant motifs}.

A {\it motif} is a small network with a particular arrangement of
internal links.  A given motif may occur many times in a network with
different rules assigned to its nodes and with different
configurations of external inputs.  A motif is {\it resistant} with
respect to a given ensemble of rule assignments if the probability of
damage entering the motif when all external inputs are damaged is
strictly less than unity.  For the rule distributions that we consider
for the \EP\ transition in random networks, each node has a nonzero
probability of being assigned a rule that sets its output to 0 if at
least one of its inputs is 0.  Thus when all of the nodes in a
feedback loop of any length have the value 0, there is a nonzero
probability that they will all remain 0 even if all external inputs to
the loop are set to 1.  Every feedback loop of a given length is
therefore a resistant motif.

If the number of occurrences of a resistant motif grows linearly with
the network size, there will in total be a finite fraction of nodes
that remain unaffected with a finite probability.  For such networks,
\EP\ cannot occur in the limit of large systems.  Examples include
typically studied regular lattices and small world networks with link
directions assigned so that short feedback loops are prevalent.

The problem resistant motifs can be avoided in random networks having
a mean indegree $\langle K\rangle$ that is well-defined and
independent of $N$, in which case the number of feedback loops of a
given length approaches a constant.  Though the total number of
resistant motifs may grow with system size, the larger motifs have a
low probability of avoiding damage.  For large $N$, the out-degree
distribution is a Poisson distribution with a mean value of $\langle
K\rangle$.  The outputs emerging from a given node form a tree with
approximately $\langle K\rangle^m$ nodes at the $m$th level.  Thus,
the probability for a given node to be part of a cycle of $m$ nodes is
approximately $\langle K\rangle^m/N$, which means that the typical
number of feedback loops of length $m$ is approximately $\langle
K\rangle^m/m$.  On the other hand, the loop may contain either
initially damaged nodes or some rules that allow damage to enter from
external inputs.  The probability that this will {\em not} occur
decays exponentially with $m$.  If the decay is faster than $\langle
K\rangle^{-m}$, the density of nodes in undamaged resistant motifs
will approach zero.

In summary, \EP\ (for the considered type of rule distributions) is excluded on
lattices with a high density of feedback loops.  For random networks,
however, the fraction of nodes in undamaged resistant motifs can go to
zero in the large $N$ limit. This property allows \EP\ to occur on
random networks as demonstrated in the following section.

\section{EP on random networks}
\label{sec: random networks}
\subsection{Criteria for EP}
\label{sec: criteria for EP}
Consider a network such that the inputs to each node are chosen
randomly and uniformly from all nodes in the network and the
probability functions $\{P_1(\kb_i)\}$ are determined from a given
distribution of Boolean rules.  For such networks, \UBA\ can be handled
analytically. 

Define $g(x)$ as the probability for a rule in the random network to
output $1$ if each input has the value $1$ with probability $x$.  The
function $g$ reflects the probability for propagation of damage to a
single node, for the considered instance of \UBA.  We refer to $g$ as
the {\it damage propagation function}.  In random networks,
$P_1(\kb_i)$ is independent of $i$ and can be replaced by $P_1(\kb)$.
Let $K$ denote the number of components of $\kb$, i.e., the number of
inputs to the considered node. $g(x)$ can then be expressed as
\begin{align}
\label{eq: g definition}
  g(x) &=
  \sum_{K=0}^\infty P(K)\!\!\!\!\sum_{\kb\in\{0,1\}^{K}}x^{I}(1-x)^{K-I}P_1(\kb),
\end{align}
where $I$ is the number of 1s in $\kb$ and $P(K)$ is the probability to
draw a rule with $K$ inputs.

Let $N$ denote the total number of nodes, and let $n_0$, $n_\msa$, and
$n_1$ denote the number of nodes with the values $0$, $\msa$, and $1$,
respectively. With these definitions and the fact that $U_i(\s,j)$ is
independent of $i$ for the random network, the role of
$\{P_1(\kb_i)\}$ is taken over by $g(n_1/N)$ and Eq.~\eqref{eq: Ui} is
replaced by
\begin{align}
  U(\s,j) &= \frac{g(n'_1/N)-g(n_1/N)}{1-g(n_1/N)},
\label{eq: n-update first}
\end{align}
where $n'_1=n_1+1$. This means that the size of the network, the
number of initially damaged nodes, and the damage propagation function
$g$ taken together are sufficient to uniquely determine the stochastic
spreading process.

After one pass of the update steps \ref{j select}--\ref{j := 1} (from
Section~\ref{sec: formal UBA}), the new values $n'_0$ and $n'_\msa$ of
$n_0$ and $n_\msa$ are given by
\begin{align}
  n'_0 &= n_0 - \delta
\intertext{and}
  n'_\msa &= n_\msa + \delta - 1
\intertext{where}
  \delta &= B_{n_0}[U(\s,j)],
\label{eq: n-update last}
\end{align}
with $B_n(a)$ being a stochastic function that returns the number of
selected items among $n$ items if the selection probability for each
of them is $a$.  The avalanche ends when $n_\msa=0$.

The number of damaged nodes, $n$, in a complete avalanche is the final
value of $n_1$, whereas the number of undamaged nodes, $u$, is the
final value of $n_0$. An order parameter for the system is $\phi =
\lim_{N\rightarrow\infty}\langle n/N\rangle$, where the average is
taken over the ensemble of networks. The \SP\ transition is found when
$\phi$ changes from zero to a nonzero value, whereas the \EP\ transition
is found when $\phi$ reaches 1.

To understand the typical development of an avalanche, it is
convenient to change from the variables $n_0$, $n_\msa$, and $n_1$,
which are constrained to sum to $N$, to the variables $x_1 \equiv
n_1/N$ and
\begin{align}
  c &\equiv \frac{n_0}{1-g(x_1)}.
\label{eq: c definition}
\end{align}
As long as $n_\msa>0$, the average value of $c$ after a single update
is given by
\begin{align}
  \langle c'\rangle 
    &= \frac{\langle n'_0\rangle}{1-g(x'_1)}\\
    &= \frac{n_0-c[g(x'_1)-g(x_1)]}{1-g(x'_1)}\\
    &= c.
\label{eq: c constant}
\end{align}
Hence, as long as $n_\msa>0$ for all members of an ensemble of
avalanches, $\langle c\rangle$ (the average of $c$ over the ensemble)
is conserved as the avalanche proceeds.

From Eqs.~\eqref{eq: n-update first}--\eqref{eq: n-update last} and
the definition of $c$, the variance in $c$ can be calculated.  We
begin by computing the increment of the variance due to one update
step, $\sigma^2(c')$.  To leading order as $N\rightarrow\infty$,
we get
\begin{align}
  \sigma^2(c')
       &= \frac{\sigma^2(\delta_0)}{[1-g(x'_1)]^2}\\
       &= \frac{n_0 U(\s,j)[1-U(\s,j)]}{[1-g(x'_1)]^2}\\
       &= \frac{c\,U(\s,j)}{1-g(x_1)}\\
       &= \frac{c}{N[1-g(x_1)]^2}\frac{dg(x)}{dx}\bigg|_{x=x_1}.
\label{eq: c-variance}
\end{align}
Eq.~\eqref{eq: c-variance} gives the increment of the variance of $c$
from one update step. To get the total variance of $c$, we need to sum
over all updates from $n_1=0$ to the desired value of $n_1$. Provided
that there is an upper bound $\kappa$ such that $dg(x)/dx<\kappa$ for all
$x$, the total variance of $c$ satisfies
\begin{align}
   \sigma_{\trm{tot}}^2(c)
       &< n_1\frac{c\,\kappa}{N[1-g(x_1)]^2}
        <\frac{\kappa N}{1-g(x_1)}
\label{eq: c tot var upper}
\end{align}
for $x_1 < 1$. (Note that $1/[1-g(x)]$ is a nondecreasing
function because $g(x)$ is nondecreasing.)

The avalanche is initiated with $n_\msa \equiv n_\msa^\ti$,
$n_0=N-n_\msa^\ti$, and $n_1=0$. The process ends when $n_0+n_1=N$ and
we seek the distribution of $n_0$ or $n_1$ when this happens.
According to Eq.~\eqref{eq: c tot var upper}, the standard deviation
of $c/N$ scales like $1/\sqrt{N}$, which implies that both $n_0/N$ and
$n_\msa/N$ have standard deviations that scale like $1/\sqrt{N}$.
($x_1$ has zero standard deviation because $n_1$ is incremented by
exactly unity on every update step.)  Thus in the large system limit,
the probability of any member of the the ensemble of avalanches
stopping is negligibly small as long as $n_\msa/N$ is finite, and we
may treat $c$ as exactly conserved as long as this condition holds.

Using the initial values $x_1 = 0$ and $n_0 = N - n_\msa^\ti$, which
determine $c$, Eq.~\eqref{eq: c definition} can be rearranged to give
\begin{align}
  n_0 &= [1-g(x_1)]\frac{N-n_\msa^\ti}{1-g(0)}.
\end{align}
Noting that $n_0/N = 1 - x_1 - n_\msa/N$, we see that
in the large $N$ limit, the process continues as long as the strict inequality
\begin{align}
  1 - x_1 &> [1-g(x_1)]
         \frac{\displaystyle 1-\lim_{N\rightarrow\infty}n_\msa^\ti/N}
              {1-g(0)}~
\label{eq: x1 fp}
\end{align}
holds, since the inequality implies that $n_\msa/N$ remains finite.
Moreover, in the large $N$ limit it is impossible to reach values of
$x_1$ for which the inequality has the opposite sign, because the
process stops when $n_\msa$ reaches zero.

Note that because of the zero probability of a node remaining
undamaged when all of its neighbors are damaged, we have $g(1)=1$,
which in turn implies that Eq.~\eqref{eq: x1 fp} becomes an equality
at $x_1 = 1$.  If Eq.~\eqref{eq: x1 fp} is satisfied for all $x_1 <
1$, the process will be exhaustive in the sense that it will not end
with a finite value of $n_0/N$.  If, on the other hand, the inequality
changes sign for $x_1$ above some threshold value, then the process
will terminate when the threshold is reached.  If the left hand side
of Eq.~\eqref{eq: x1 fp} forms a tangent line to the right hand side
of the expression at some value of $x_1$, the process will exhibit
critical scaling laws.  The critical case for \EP\ occurs when the
when the tangency occurs at $x_1 = 1$.  Examples of these behaviors
are presented below and in Section~\ref{sec: application}.

As an aside, we note that the \SP\ transition is an instance of
criticality at $x_1=0$. For the above mentioned criterion of
criticality to hold at $x_1=0$, the right hand
side of Eq.~\eqref{eq: x1 fp} must have the value $1$ and the slope $-1$ at
$x_1=0$. Thus, the system is critical with respect to \SP\ if
$\lim_{N\rightarrow\infty}n_\msa^\ti/N=0$ and
\begin{align}
\label{eq: critical g}
  \frac{dg(x)}{dx}\bigg|_{x=0} &= 1 - g(0).
\end{align}

Eqs.~\eqref{eq: g definition} and~\eqref{eq: critical g} immediately
give a criterion for critical percolation on graphs in which every
possible directed link (including self-inputs) exists with an
independent, fixed probability, assuming the conventional choice in
which damage spreads to a given node with probability $p$ from each of
its damaged neighbors.  In this case we have
\begin{align}
 g(x) &= 
 \sum_{K=0}^\infty P(K)\bigl[1 - (1-px)^K\bigr],
\end{align}
which yields
\begin{align}
  \frac{dg(x)}{dx}\bigg|_{x=0} &= p \sum_{K=0}^\infty P(K) K \\
  \ &= p \langle K \rangle.
\label{eq: ER percolation}
\end{align}

This result is closely related to the well-known criterion for the
presence of a percolating cluster in an Erd\H{o}s--R\'{e}nyi graph:
percolation occurs when the probability $p_{\trm {\sc er}}$ for the
presence of a link between two randomly selected nodes exceeds $1/N$,
where $N$ is the number of nodes.~\cite{Bollabas:85} In the present
context, $p_{\trm {\sc er}}$ is mapped to $p_{\trm{link}} p$, where
$p_{\trm{link}}$ is the probability that a link exists connecting the
two randomly selected nodes and $p$ is the probability that damage
spreads across that link.  At the same time, we have $\langle K\rangle
= p_{\trm{link}} N$.  (Recall that $K$ is only the indegree of a node,
not the total number of links connected to it.)  Thus Eq.~\eqref{eq:
  ER percolation}, which implies that the critical value of $p$ is
$1/\langle K \rangle$, is consistent with the well-known theory of
Erd\H{o}s--R\'{e}nyi graphs.~\cite{Bollabas:85}

Eq.~\eqref{eq: ER percolation} applies for any distribution of
indegrees so long as $\langle K \rangle$ is well-defined and the
source of each input is selected at random (so that the outdegrees are
Poisson distributed).  We note that the latter condition is {\em not}
met by random regular graphs (graphs in which all nodes have the same
outdegree) because the probabilities of two nodes getting an output
from the same node are correlated.

\SP\ can also be understood by the theory of Galton--Watson processes.
If $\lim_{N\rightarrow\infty}n_\msa^\ti/N=0$, the update described by
Eqs.~\eqref{eq: n-update first}--\eqref{eq: n-update last} is
consistent with a Galton--Watson processes that has a Poisson
out-degree distribution with a mean value
\begin{align}
  \lambda &= \frac1{1 - g(0)}\frac{dg(x)}{dx}\bigg|_{x=0}.
\end{align}
See References~\cite{Harris:63, Otter:49, Ramo:06}. See
Appendix~\ref{app: SP} for more details on \SP\ in relation to known
results. Cases of tangencies at intermediate values of $x_1$ are
beyond the scope of the present work.

Returning to the question of the \EP\ transition, it is convenient to
change variables once again.  We define $x_\msia \equiv 1-x_1$ and
$q(x_\msia) \equiv 1-g(x_1)$.  In words, $q(x)$ is the probability
that a randomly selected node will output 0 given that each of its
inputs has the value 0 with probability $x$.  We refer to $q$ as the
{\it damage control function} as it characterizes the probability that
damage will be prevented from spreading to a single node.
Equation~\eqref{eq: x1 fp} is then transformed to
\begin{align}
  x_\msia &> q(x_\msia)
           \frac{\displaystyle 1-\lim_{N\rightarrow\infty}n_\msa^\ti/N}
                {q(1)}.
\label{eq: x0 gt}
\end{align}
Critical \EP\ is found when the left hand side of Eq.~\eqref{eq: x0
  gt} forms a tangent line to the right hand side of the expression at
$x_\msia=0$.  At criticality, the right hand side of Eq.~\eqref{eq: x0
  gt} should have the value 0 and the slope 1.  Hence, the conditions
$q(0)=0$ and
\begin{align}
  \frac{dq(x)}{dx}\bigg|_{x=0} &= \frac{q(1)}{1-n_\msa^\ti/N}
\label{eq: EP criticality}
\end{align}
are required for an \EP\ transition.

\subsubsection*{Example: EP on random digraphs}
We now consider the special case of graphs in which every possible
directed link (including self-inputs) exists with an independent,
fixed probability.  (We have already discussed \SP\ on such graphs.)
If damage spreads along each directed link with probability $p$, there
is no \EP\ transition because there is a nonzero probability for a
node to remain undamaged when all of its inputs are damaged.  A
minimal change that allows \EP\ on such graphs is to give a special
treatment to nodes whose inputs are all damaged, in which case the
considered node should always get damaged.  For the same reason, all
nodes with no inputs must be initially damaged.  Other nodes might
also be initially damaged, and we let this happen with a given
probability $\rho$ for each node with at least one input.  For such a
network, we can calculate the damage propagation function according to
\begin{align}
  g(x) &= \sum_{K=0}^\infty P(K)
  \bigl[1-(1-px)^K+(1-p)^Kx^K\bigr]\\
  &= 1-e^{-\langle K\rangle px}\bigl(1-e^{-\langle
    K\rangle(1-x)}\bigr).
\end{align}

The corresponding damage control function becomes
\begin{align}
  q(x) &= e^{-\langle K\rangle p(1-x)}\bigl(1-e^{-\langle K\rangle x}\bigr).
\end{align}
A necessary condition for the \EP\ transition is derived from
Eq.~\eqref{eq: EP criticality}, yielding
\begin{align}
  \langle K\rangle e^{-p\langle K\rangle} &= \frac1{1-\rho}~.
\label{eq: E-R EP slope}
\end{align}
For the \EP\ transition to occur, it is also required that
\begin{align}
  f(x) \equiv x - q(x)(1-\rho) &\ge 0
\label{eq: E-R EP ineq}
\end{align}
for all $x\in[0,1]$ according to Eq.~\eqref{eq: x0 gt}. If both
Eqs.~\eqref{eq: E-R EP slope} and~\eqref{eq: E-R EP ineq} are
satisfied, the \EP\ transition occurs at the value of $p$
given by Eq.~\eqref{eq: E-R EP slope}:
\begin{align}
  p_{\trm c} &= \frac{\ln\langle K\rangle+\ln(1-\rho)}{\langle K\rangle}~.
\end{align}

Equation~\eqref{eq: E-R EP slope} turns out to be a sufficient and
necessary condition for the \EP\ transition.  Provided that
Eq.~\eqref{eq: E-R EP slope} holds, the first derivative satisfies
$f'(0)=0$.  From the observation $f'''(x)<0$, it is then
straightforward to show that $f(x)$ has no local minimum on the
interval $(0,1)$.  Since $f(0)=0$ and $f(1)>0$, Eq.~\eqref{eq: E-R EP
  ineq} holds for all $x\in[0,1]$.

\begin{figure}[bt]
\begin{center}
\includegraphics{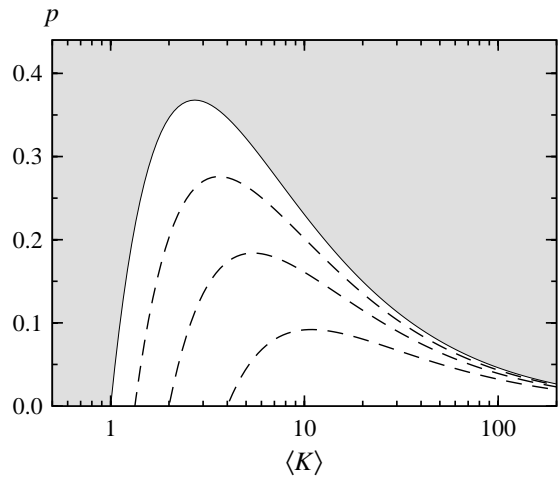}
\end{center}
\caption{\label{fig: E-R EP phase} Phase diagram for \EP\ on 
  random digraphs, where damage spreads along each directed link with
  probability $p$ and a node is guaranteed to get damaged in the
  special case that all of its inputs are connected to damaged nodes.
  All nodes with zero inputs are initially damaged, and the other
  nodes are initially damaged with probability $\rho$.  The gray area
  bounded by a solid line shows the region where \EP\ occurs for
  $\rho=0$ and the dashed lines show the \EP\ transition when $\rho$
  has the values $1/4$, $1/2$, and $3/4$, respectively.}
\end{figure}

It is instructive to examine the phase diagram at fixed $\rho$.  A
negative value of $p_{\trm c}$ indicates that the system is always in
the \EP\ regime, so for $\langle K\rangle < 1$ the system exhibits
\EP\ and it is not possible to observe a transition.  For $\langle
K\rangle > 1$, an \EP\ transition can be observed at $p=p_{\trm c}$.
A curious feature of this system is that $p_{\trm c}$ is not a
monotonic function of $\langle K\rangle$, having a maximum value of
$(1-\rho)/e$ at $\langle K\rangle = e/(1-\rho)$ and approaching zero
as $\langle K\rangle$ approaches infinity.  Thus if $p$ is held fixed
at any value between zero and $(1-\rho)/e$, the system will undergo
two transitions as $\langle K\rangle$ is increased from zero.  The
system will begin in the \EP\ regime (i.e. $p>p_{\trm c}$), undergo a
transition to subcritical behavior at some $\langle K\rangle$, then
reenter the \EP\ regime for a higher value of $\langle K\rangle$.  The
calculated phase diagram is shown in Fig.~\ref{fig: E-R EP phase} and
has been verified by direct numerical simulations of avalanches.
Roughly speaking, at low $\langle K\rangle$ \EP\ occurs due to the
high density of initially damaged nodes with no inputs.  At high
$\langle K\rangle$, on the other hand, \EP\ occurs due to the high
probability of nodes being damaged because of their large number of
inputs.

\subsection{The probability of complete coverage}
\label{sec: cc}
An important quantity associated with \EP\ is the probability of an
avalanche yielding complete coverage of the system; i.e., the
probability that all sites are damaged by the \UBA\ so that $u=0$.
Let $\Pex(N,q; n_0,n_\msa)$ denote the probability that a \UBA\ on a
random network will yield complete coverage for a system with a given
network size $N$, a given damage control function $q$, and starting
with particular values of $n_0$ and $n_\msa$.  For future convenience
we also define $\Pex(N,q)$ to be the probability for complete coverage
assuming that each node is initially damaged with probability $1-q(1)$
and we average over the corresponding probability distribution for
$n_\msa$.

To calculate $\Pex(N,q; n_0,n_\msa)$, we note that 
\begin{align}
\Pex(N,q;m,0)=0\quad {\rm if\ } m>0,
\end{align} 
since the process stops when $n_\msa = 0$.  We also have 
\begin{align}
  \Pex(N,q;0,m)=1\quad {\rm for\ any\ } m,
\end{align} 
since updating can never create 0s.  These values of $\Pex$ can be
used for recursive calculation of $\Pex$. Let $n_\msia$ denote
$n_0+n_\msa$, or $Nx_\msia$.  Performing steps \ref{j select}--\ref{j
  := 1} (from Section~\ref{sec: formal UBA}) one time decreases
$n_\msia$ by $1$ as described by Eqs.~\eqref{eq: n-update
  first}--\eqref{eq: n-update last}.  This means that $\Pex(N,q;
n_0,n_\msa)$ can be calculated for all $n_\msia=m$ if $\Pex(N,q;
n_0,n_\msa)$ is known for all $n_\msia=m-1$. The recursion starts at
$n_\msia=0$ with $\Pex(N,q;0,0)=1$ and uses the boundary conditions
$\Pex(N,q; n_\msia,0)=0$ and $\Pex(N,q; 0,n_\msia)=1$ for $n_\msia>0$.

For large $N$, $\Pex$ can be calculated in the framework of a
continuous approximation. Let $p(n_\msia,c)$ denote a continuous version
of $\Pex(N,q$; $n_0,n_\msa)$. Then, the boundary conditions
$\Pex(N,q; n_\msia,0)=0$ and $\Pex(N,q$; $0,n_\msia)=1$ are
expressed as
\begin{align}
  p[n_\msia,c_{\trm{max}}(x_\msia)] &= 0,
\label{eq: p-diff bound beg}
\intertext{and}
  p(n_\msia,0) &= 1,
\intertext{where}
  c_{\trm{max}}(x_\msia) &= \frac{n_\msia}{q(x_\msia)}.
\label{eq: p-diff bound end}
\end{align}

In the continuous approximation, the recurrence relation that can be
derived from Eqs.~\eqref{eq: n-update first}--\eqref{eq: n-update
last} is transformed to a partial differential equation.  In such an
update, the change $n_\msia$ decreases by unity and, for large $N$,
the change in $c$ is much less than $c$ itself.  In the continuous
approximation, this means that $p(n_\msia,c)$ satisfies a partial
differential equation of the form
\begin{align}
  \frac{\partial p}{\partial n_\msia} 
        &= h_1(n_\msia,c)\frac{\partial p}{\partial c}
           + h_2(n_\msia,c)\frac{\partial^2p}{\partial c^2},
\label{eq: p-diff 0}
\end{align}
where $h_1(n_\msia,c)$ and $h_2(n_\msia,c)$ are functions to be
determined.  This is recognizable as a 1D diffusion equation in which
$n_\msia$ plays the role of time and $c$ the role of space.  Note that
later times in the diffusion equation correspond to earlier stages of
the \UBA, since $n_\msia$ decreases as nodes are converted to 1s.  The
boundary conditions on the diffusion are given by Eqs.~\eqref{eq:
  p-diff bound beg} and~\eqref{eq: p-diff bound end}.  We are
interested in computing $p(n_{\msia},c)$ for values of $n_\msia$ and
$c$ corresponding to $n_\msa = n_\msa^\ti$ and $n_1 = 0$.

The fact that the average of $c$ is constant means that the
coefficient of the drift term in the diffusion equation must vanish;
i.e., $h_1(n_\msia,c)=0$.  The diffusion coefficient,
$h_2(n_\msia,c)$, is given by
\begin{align}
  h_2 &= \tfrac12\sigma^2(c'),
\label{eq: h2}
\end{align}
where $\sigma^2(c')$ is the variance of $c'$ when a fixed $c$ is
updated.  

Using Eqs.\ \eqref{eq: c-variance} and \eqref{eq: h2} and converting
$g$'s to $q$'s, we find
\begin{align}
  \frac{\partial p}{\partial n_\msia} 
        &= \frac{c}{2N[q(x_\msia)]^2}
           \frac{dq(x)}{dx}\bigg|_{x=x_\msia}
           \frac{\partial^2p}{\partial c^2}.
\label{eq: p-diff 1}
\end{align}

The large $N$ behavior of Eq.~\eqref{eq: p-diff 1}, with the boundary
conditions in Eqs.\ \eqref{eq: p-diff bound beg} and \eqref{eq: p-diff
bound end}, can be found by expanding $q(x)$ around $x=0$. If $q(x)$
is well-behaved, such an expansion can be written as
\begin{align}
  q(x) &= \alpha_1x - \alpha_2x^2 + \Oc(x^3).
\label{eq: q-Taylor}
\end{align}
This expansion can always be performed if the probability $P(K)$ for a
node to have $K$ inputs decays as least as fast as $K^{-4}$ and in the
case that $p_K$ decays slower than $K^{-4}$ but faster than $K^{-3}$,
only the residue term can be affected.  See Appendix \ref{app: q
  calc}.  In particular, the expansion is always valid if $K$ has a
maximal value.

The most interesting case in terms of asymptotic behavior is when
$\alpha_1$ is close to $1$ and $\alpha_2$ is positive. With suitable
$N$-dependent transformations of $p$ and its arguments, described in
Appendix \ref{app: PEP}, the large $N$ behavior of Eq.~\eqref{eq:
  p-diff 1} can be expressed in terms of a function $\pt(\tti,\yt)$
determined by the differential equation
\begin{align}
  \frac{\partial\pt}{\partial\tti} &= \frac12
         \frac{\partial^2\pt}{\partial\yt^2},
\label{eq: diffuse}
\end{align}
with the boundary conditions
\begin{align}
  \pt(\tti,1/\tti) &= 0 \quad\trm{for }\tti<0
\label{eq: diffuse bound 0}
\intertext{and}
  \lim_{\tti\rightarrow-\infty}\pt(\tti,\yt) &= \yt
           \quad \trm{for }\yt\ge0.
\label{eq: diffuse bound 1}
\end{align}
The Crank--Nicholson method can be used to calculate $\pt(\tti,\yt)$
numerically in an efficient way. (See, e.g., \cite{numrecip}.)

Appendix~\ref{app: PEP} shows that the probability for complete
coverage is given by
\begin{align}
  \Pex(N, q) &\approx \Nt^{-1/3}\pt[0,\Nt^{1/3}(1-\alpha_1)],
\label{eq: PexNq}
\end{align}
where $\Nt=\alpha_1N/\alpha_2$.  The calculation assumes that the
avalanche is initiated on the nodes whose outputs are independent of
their inputs, as accounted for in $q(1)$.

To our knowledge, the critical point for \EP\ has not been
investigated previously in its own right.  Two special cases have been
studied, however.  First, results for numbers of frozen and unfrozen
nodes in critical \RBN s can be mapped to an \EP\ process, as
discussed in Section~\ref{sec: application}.  In this context, frozen
nodes in the network are considered to be the damaged nodes of the
\UBA, and the scaling with $N$ of the number of unfrozen nodes at the
phase transition has been investigated for certain class of \RBN s
\cite{Socolar:03,Kaufman:05}.

Second, in the special case that $q(x)=x$, the exact result
\begin{align}
   \Pex(N,x\mapsto x;n_0,n_\msa)
             &= \frac{n_\msa}{n_0+n_\msa}
\label{eq: q-ident main}
\end{align}
is obtained.  [See Eq.~\eqref{eq: q-ident} in Appendix \ref{app:
  PEP}.] This means that the probability for complete coverage is
exactly $n^\ti_\msa/N$. The simplest realization of $q(x)=x$ is
provided by a network of one-input nodes with rules that copy the
input state.  Such networks have strong connections to random maps
from a set of $N$ elements into itself.  A map $T$ is derived from a
network of one-input nodes by letting each node map to the node from
where its input is taken. In this picture, the damage originating from
one initially damaged node $i$, corresponds to the set of nodes $j$
such that $T^k(j)=i$ for some $k\ge0$ (where $T^k$ denotes the $k$th
iterate of $T$). Such a $j$ is called a {\it predecessor} to $i$. See,
e.g., Ref.~\cite{Bollabas:85} for an overview of the theory of
random maps and see Refs.~\cite{Rubin:54, Harris:60} for results on
predecessors in random maps. See Appendix~\ref{app: exact} for
analytic results that relate \UBA\ to random maps.

\subsection{On the number of damaged nodes in random networks}
\label{sec: avalanche size}

In the Sections~\ref{sec: criteria for EP} and~\ref{sec: cc} we
focused on determining the parameters that lead to \EP\ (a vanishing
fraction of undamaged nodes large $N$ limit) and on the probability
that the number of undamaged nodes will be exactly zero (complete
coverage).  We now consider the full probability distribution for the
number of nodes damaged in an avalanche in a manner that provides a
suitable base for understanding both \SP\ and \EP\ in random networks.
The calculational strategy involves considering a given set of $n$
nodes to be the damaged set and computing the probability that this is
both consistent with all of the Boolean rules and the probability that
the avalanche will actually cover the whole set.  The probability of
consistency is calculated via elementary combinatorics.  The
probability of reaching the whole set is precisely the probability of
complete coverage for an avalanche on the sub-network of $n$ candidate
nodes.  For this we can directly apply the results of the last
section.  For the purposes of explaining the calculation, we refer to
the selected set of $n$ nodes as the {\it candidate set}.

We let $\Pn(n)$ denote the probability that $n$ nodes will be damaged
in an avalanche, averaged over the ensemble of $N$-node networks with
a rule distribution characterized by a given damage propagation
function $g$ or the corresponding damage control function $q$.  We
assume that the avalanche is initiated by randomly selecting $\ell$
nodes to set to $\msa$, regardless of their Boolean rules, then
setting to $\msa$ all nodes with rules that always output 1 for any
inputs.  The set of $\ell$ initially damaged nodes must be a subset of
the candidate set.  The probability that the candidate set contains
all of the nodes with ``always 1'' rules will be taken into account by
the value of $g(0)$ in the expression below for the consistency
probability.  We use the notation $\binom m k$ for the usual binomial
coefficient (the number of combinations of $k$ objects chosen from a
set of $m$ objects).

The probability $\Pn(n)$ can be expressed as
\begin{align}
\Pn(n) = \binom{N-\ell}{n-\ell} P_{\trm c}(n, \ell; N)\, \PNex(n,\ell;N),
\label{eq: Pnn}
\end{align}
where $P_{\trm c}(n, \ell; N)$ and $\PNex(n,\ell;N)$ are defined below.
The binomial factor counts the number of different sets of $n-\ell$
nodes that could be damaged in a process corresponding to a given set
of $\ell$ nodes that are initially damaged without regard to their
rules.  $P_{\trm c}(n, \ell; N)$ is the probability that a given choice of
$n-\ell$ nodes assumed to be damaged by the avalanche will constitute
a final state that is consistent with the Boolean rules for each node,
including the nodes that are initially damaged because their rules
require it.  $\PNex(n,\ell;N)$ is the probability that the avalanche
will not die out before damaging all $n$ nodes.  This factor is
necessary to avoid counting final states that contain loops of damaged
nodes consistent with the rules but unreachable because damage cannot
spread to the loop from any nodes outside the loop.

Consistency with the Boolean rules requires that the given set of
$n-\ell$ nodes damaged in the avalanche have inputs that cause them to
be damaged.  In a random network, the probability that any single node
will be damaged is $g(x_1)$, where $x_1$ is the fraction of damaged
nodes.  Similarly, the probability that any node will {\it not} be
damaged is $1-g(x_1)$.  We are considering candidate sets of damaged
nodes with $x_1 = n/N$.  Thus we have
\begin{align}
  P_{\trm c}(n,\ell;N) = [g(n/N)]^{n-\ell}[1-g(n/N)]^{N-n}.
\label{eq: Pconsistency}
\end{align}

The computation of $\PNex(n,\ell;N)$ involves the rule distribution on
the restricted network formed by the candidate set with all inputs
from the undamaged nodes removed.  This distribution, $g^1(x)$, is
different from $g(x)$ because $P_{\trm c}$ already accounts for rules that
are not consistent with the pattern of damage.  Thus the spreading of
damage on the $n$-node network involves $g(nx/N)$, the probability
that a rule outputs $1$ when a fraction $x$ of the $n$-node candidate
set is damaged.  The probability must be normalized such that it goes
to unity when $x$ goes to 1.  (We know that a node in the $n$-node set
should get damaged if all of its inputs are damaged.)  Thus
we have
\begin{align}
g^1_{N,n}(x) = \frac{g(nx/N)}{g(n/N)}
\label{eq: g1}
\end{align}
or, equivalently, 
\begin{align}
  q^1_{N,u}(x) &= \frac{q[u/N+(1-u/N)x]-q(u/N)}{1-q(u/N)}.
\label{eq: q1}
\end{align}
(Recall that $u = N-n$ is the number of undamaged nodes after an
avalanche.)

There are two cases of interest for the probability of complete
coverage of the candidate set.  For \EP, $g(0)>0$ and the fixed number
$\ell$ of nodes arbitrarily selected for damage is irrelevant compared
to the finite fraction of nodes with rules that produce damage for any
combination of inputs.  In this case, we assume $\ell=0$, which allows
reduction of $\Pex$ to its two-argument form defined at the beginning of
Section~\ref{sec: cc}:
\begin{align}
  \PNex(n,0;N) & = \Pex(n,q^1_{N,N-n}).
\label{eq: cc on n EP}
\end{align}

For \SP, we have $g(0)=0$ so the avalanche must be initiated with a 
nonzero value of $\ell$.  In this case we have
\begin{align}
  \PNex(n,\ell;N) = \Pex(n,q^1_{N,N-n};n-\ell,\ell).
\label{eq: cc on n SP}
\end{align}
Note that $\PNex(n,\ell;N)$ depends on $N$ only through $q^1$.

For notational convenience, we now let $\PNex$ stand for whichever
expression on the right-hand side of Eqs.~\eqref{eq: cc on n EP}
or~\eqref{eq: cc on n SP} is relevant, and we use $u$ where $N-n$
would be the strictly proper form. By combining Eqs.~\eqref{eq: Pnn}
and \eqref{eq: Pconsistency}, we get
\begin{align}
  \Pn(n) =\,&\binom{N-\ell}{n-\ell}[g(n/N)]^{n-\ell}[1-g(n/N)]^{u}
           \PNex.
\label{eq: Pn0}
\end{align}

To make some important features of Eq.~\eqref{eq: Pn0} apparent, we
introduce the functions
\begin{align}
  \rho(n) &= \frac{n^n}{e^nn!},\\
  \tau(n, k) &= \frac{n!}{n^k(n-k)!},
\intertext{and}
  G(x) &= \biggl(\frac{g(x)}x\biggr)^{\!x}
          \biggl(\frac{1-g(x)}{1-x}\biggr)^{\!1-x}.
\end{align}
Then Eq.~\eqref{eq: Pn0} can be rewritten as
\begin{align}
  \Pn(n) 
      =\,&\frac{\rho(n)\rho(u)}{\rho(N)}
          \frac{\tau(n, \ell)}{\tau(N, \ell)}
          \biggl(\frac{n/N}{g(n/N)}\biggr)^{\!\ell} \nonumber\\ 
         &\times [G(n/N)]^N \PNex.
\label{eq: Pn1}
\end{align}

Stirling's formula,
\begin{align}
  n! &\approx \sqrt{2\pi n}\,\frac{n^n}{e^n},
\label{eq: Stirling}
\end{align}
yields
\begin{align}
  \rho(n)&\approx\frac1{\sqrt{2\pi n}}
\intertext{and}
  \frac{\rho(n)\rho(u)}{\rho(N)}&\approx\frac1{\sqrt{2\pi nu/N}}.
\end{align}

The factor $\tau(n,\ell)/\tau(N,\ell)$ is approximately 1 for large $n$ and
satisfies
\begin{align}
  \frac{\tau(n, \ell)}{\tau(N, \ell)}&\le1
\end{align}
for $n\leq N$, with equality if $n=N$ or $\ell=1$ or $\ell=0$. The
only factors in Eq.~\eqref{eq: Pn1} that can show exponential
dependence on $N$ are the $G$ and $\PNex$ factors.  Because $\PNex$ is
a probability (and therefore cannot exceed unity) and $G(x)\leq1$ with
equality if and only if $g(x)=x$, $\Pn(n)$ vanishes exponentially as
$N$ goes to infinity for any fixed $n/N$ such that $g(n/N)\ne n/N$.
This is consistent with the above result that the probability of an
avalanche stopping with $x_1 \neq g(x_1)$ is vanishingly small.  [See
Eqs.~\eqref{eq: x1 fp} and~\eqref{eq: x0 gt}.]

For \EP, we are interested in the number of undamaged nodes, $u$. We
let
\begin{align}
  \Pu(u) &= \Pn(N-u)
\intertext{and}
  Q(x) &= G(1-x) \nonumber \\
       &= \biggl(\frac{1-q(x)}{1-x}\biggr)^{\!1-x}
          \biggl(\frac{q(x)}x\biggr)^{\!x}.
\end{align}
For \EP, $g(0)>0$ and a fixed $\ell$ is irrelevant when
$N\rightarrow\infty$. Hence, we let $\ell=0$ and rewrite
Eq.~\eqref{eq: Pn1} as
\begin{align}
  \Pu(u) =\frac{\rho(n)\rho(u)}{\rho(N)} [Q(u/N)]^N \PNex.
\label{eq: Pu0}
\end{align}

To some respects, $\Pu$ is similar to $\Pn$: the factor
$[\rho(u)\rho(n)]/\rho(N)$ is fully symmetric with respect to
interchange of $n$ and $u$; and the role of $G(n/N)$ in Eq.~\eqref{eq:
  Pn1} is identical to the role of $Q(u/N)$ in Eq.~\eqref{eq: Pu0}.
However, the behavior of $\PNex$ for $n\ll N$ given by Eq.~\eqref{eq:
  cc on n SP} is significantly different from the behavior of $\PNex$
for $u\ll N$ given by Eq.~\eqref{eq: cc on n EP}.

For \EP, we consider damage control functions $q(x)$ that can be
expanded according to Eq.~\eqref{eq: q-Taylor}. For supercritical \EP,
with $\alpha_1<1$, $\Pu(u)$ decays exponentially with $u$. In Appendix
\ref{app: super EP}, we demonstrate that
\begin{align}
  \lim_{N\rightarrow\infty}\Pu(u) 
        &= (1-\alpha_1)\frac{(u\alpha_1)^{u}}{u!}e^{-u\alpha_1}
\label{eq: pu(u) exact asympt}
\\ 
         \ &\approx\frac{1-\alpha_1}{\sqrt{2\pi}}e^{u(1-\alpha_1)}
            \alpha_1^uu^{-1/2}.
\label{eq: pu(u) asympt}
\end{align}

For critical \EP, Eq.~\eqref{eq: PexNq} gives
\begin{align}
  \PNex(n,0;N) & = \Pex(n,q^1_{N,N-n}) \nonumber \\
               & \approx \tilde{n}^{-1/3}\pt[0,\tilde{n}^{1/3}(1-\alpha_1^1)],
\label{eq: cc on n crit EP}
\end{align}
where $\tilde{n} \equiv \alpha_1^1 n/\alpha_2^1$ and $\alpha_1^1$ and
$\alpha_2^1$ are the first two coefficients of the power series
expansion of $q^1(x)$ about $x=0$.  With $\alpha_1=1$ and
$\alpha_2>0$, a Taylor expansion of $\log Q(x)$ about $x=0$ gives
\begin{align}
  Q(x) &\approx \exp\biggl(-\frac{\alpha_2^2x^3}2\biggr)
\end{align}
for small $x$. This yields that the typical number of undamaged nodes,
$u$, scales like $N^{2/3}$. In Appendix \ref{app: crit EP}, we derive
the asymptotic distribution of $u$ for large $N$. With
$\ut=\Nt^{-2/3}u = (\alpha_2/N)^{2/3}u$, we find that the large $N$
limit of the probability density for $\ut$ is
\begin{align}
  P(\ut) &= \frac{\exp(-\frac12\ut^3)}
                  {\sqrt{2\pi\ut}}\,\pt(0,2\ut).
\label{eq: p(ut) asympt}
\end{align}

Eq.~\eqref{eq: Pn1} is suitable for understanding \SP\ as well as \EP.
For \SP, $g(0)=0$ and $\ell>0$. In the large $N$ limit, \SP\ is a
branching process with a Poisson distribution in the number of
branches from each node. The average number of branches per node is
given by the derivative of $g(x)$ at $x=0$, because
$\lim_{x\rightarrow0} g(x)/x$ is the average number of nodes that will
be damaged in one update as a direct consequence of damaging a single
node in the large network limit.  In Appendix \ref{app: SP}, we
re-derive known results on \SP\ in the framework of our formalism.

\section{An application:  Frozen nodes in random Boolean networks}
\label{sec: application}

An important application of our results on \EP\ in random networks is
the determination of the size distribution for the set of unfrozen
nodes in 2-input random Boolean networks, a subject of interest since
the introduction of the Kauffman model in 1969 \cite{Kauffman:69}.
The Kauffman model was originally proposed as a vehicle for studying
aspects of the complex dynamics of transcriptional networks within
cells.

In a Boolean network, there are usually some nodes that will reach a
fixed final state after a transient time regardless of the initial
state of the network.  For most random Boolean networks, nearly all of
these nodes can be found by a procedure introduced in
Ref.~\cite{Flyvbjerg:88b} and applied numerically in
Ref.~\cite{Bilke:01}.  We refer to nodes identified by this procedure
as {\it frozen}.

The nodes that cannot be identified as frozen are labeled {\it
  unfrozen}.  Their output may switch on and off for all time or
simply have different values on different attractors of the network
dynamics.  A frozen node will always reach its fixed final state
regardless of the initial state of the network.  The converse is not
true: an unfrozen node can have a fixed final state that is
independent of the initial state due to correlations that are not
accounted for in the identification procedure for frozen nodes.  In a
typical random Boolean network, the number of nodes that are
mislabeled in this sense is negligible \cite{Bilke:01}.  For the
purposes of investigating dynamics of the network at long times, one
is interested in the size of the unfrozen set.

The procedure for identification of the frozen nodes starts by marking
all nodes with a constant output function as frozen.  There may then
be nodes that, as a consequence of receiving one or many inputs from
frozen nodes, will also produce a constant output.  These nodes are
also marked as frozen, and the process continues iteratively until
there are no further nodes that can be identified as frozen.

We note here that the process of finding frozen nodes in a \RBN\ can
often be framed as a \UBA, where the property of being frozen
corresponds to damage.  That is, the process of identifying frozen
nodes involves continually checking all nodes to see whether their
inputs are frozen in such a way that they themselves become frozen, a
process which satisfies the conditions for \UBA.  The damage
propagation and damage control functions for the \UBA\ are determined
by the relative weights of different Boolean logic functions in the
\RBN.  By changing these weights, one can observe a transition in the
dynamical behavior of \RBN s corresponding precisely to the \EP\ 
transition in the \UBA.  We consider here \RBN s with exactly two
inputs at each node, with some explicit choices for the weights of the
Boolean logic functions that permit observation of both sides of the
transition.

The only restriction required for mapping the freezing of nodes in a
\RBN\ to a \UBA\ system is that the logic functions in the \RBN\ be
symmetric with respect to the probability of freezing being due to
{\sc true} and {\sc false} inputs.  That is, the probability that a
node with a certain set of frozen inputs will itself be frozen should
not depend on the values of the frozen inputs.  This condition is
satisfied for the most commonly investigated classes of rule
distributions, where there is a given probability $p$ for obtaining a
1 at each entry in the truth table for each rule.  If the above
mentioned symmetry condition were violated, it would be necessary to
distinguish nodes frozen {\sc true} from nodes frozen {\sc false},
which would mean that the state of a node could not be specified by a
binary variable.  For the rest of this section we consider only \RBN s
that respect the symmetry condition.

It is useful to distinguish different types of Boolean logic
functions.  A {\it canalizing} rule is one for which the output is
independent of one of the inputs for at least one value of the other
input.  Among the 16 possible 2-input Boolean rules, 2 rules are
constant (``always on'' or ``always off''), 12 rules are non-constant
and canalizing, and 2 rules are non-canalizing ({\sc xor} and not-{\sc
  xor}).  The original version of the Kauffman model assumes that all
2-input Boolean rules are equally likely, which turns out to give
critical dynamics.

Let $p_i$ denote the probability that a randomly selected node's
output is frozen if exactly $i$ of its inputs are frozen.  The damage
propagation function $g(x)$ can be expressed directly in terms of
$p_i$:
\begin{align}
  g(x) = p_0(1-x)^2 + 2p_1x(1-x) + p_2x^2.
\end{align}
Nodes with constant rules are guaranteed to be frozen.  (These nodes
will initiate the \UBA.)  Nodes with non-constant canalizing rules are
unfrozen if both inputs are unfrozen, and they are frozen with
probability $1/2$ if exactly one randomly selected input is frozen.
Nodes with rules that are non-canalizing become frozen if and only if
both of their inputs are frozen.  Finally, if both inputs are frozen,
the output of any 2-input rule is frozen.  Thus for the 2-input
Kauffman model, $p_0=1/8$, $p_1=1/2$, and $p_2=1$.

\begin{figure}[bt]
\begin{center}
\includegraphics{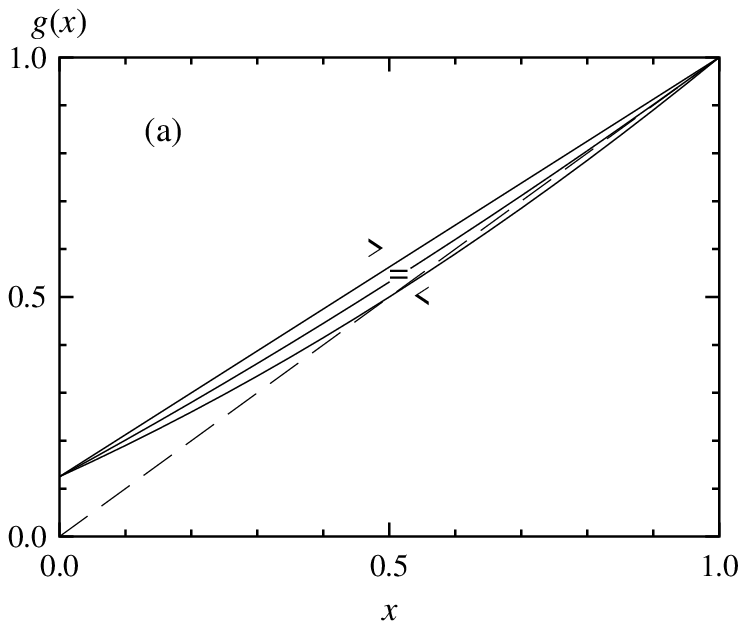}
\includegraphics{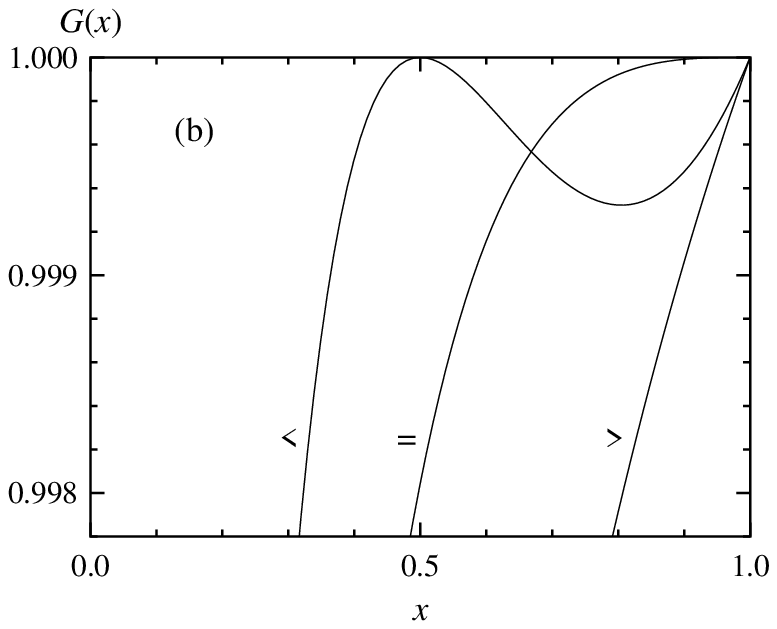}
\end{center}
\caption{\label{fig: gq} The functions (a) $g(x)\equiv 1-q(1-x)$ and
  (b) $G(x)\equiv Q(1-x)$ for three 2-inputs rule distributions. All
  three distributions have $p_0=1/8$ and $p_2=1$, whereas $p_1$ takes
  the values $7/16$, $1/2$, and $9/16$. The case that has $p_1=1/2$
  (marked with $=$) is critical with respect to \EP\ and corresponds
  to the propagation of frozen node values in the original Kauffman
  model. The other cases $p_1=7/16$ ($<$) and $p_1=9/16$ ($>$) are
  subcritical and supercritical, respectively.  The dashed line in (a)
  shows the identity function $x\mapsto x$.}
\end{figure}

If the two non-canalizing rules in the 2-input Kauffman model are
replaced by canalizing rules, $p_1$ becomes $9/16$, whereas $p_0$ and
$p_2$ are unchanged. Such networks exhibit supercritical \EP. To get a
subcritical network, we replace two of the canalizing rules with
non-canalizing rules and get $p_1=7/16$.  (Note that some care must be
taken to maintain the {\sc true}--{\sc false} symmetry mentioned
above.)  The functions $g(x)$ and $G(x)$ for critical, supercritical,
and subcritical rule distributions are shown in Fig.~\ref{fig: gq}.

\begin{figure}[bt]
\begin{center}
\includegraphics{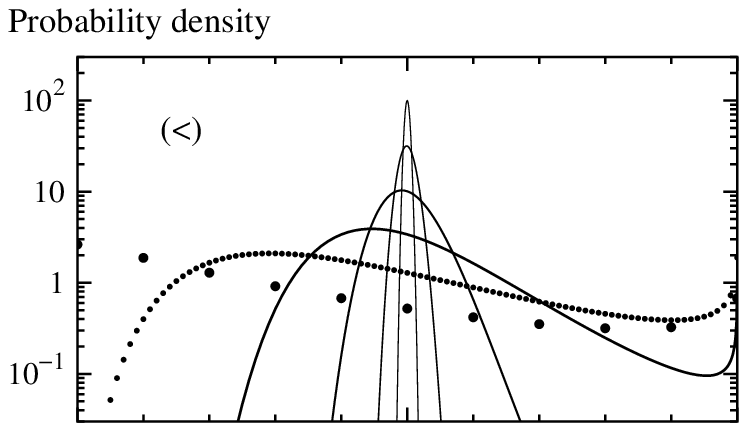}
\\\vspace*{-42pt}
\includegraphics{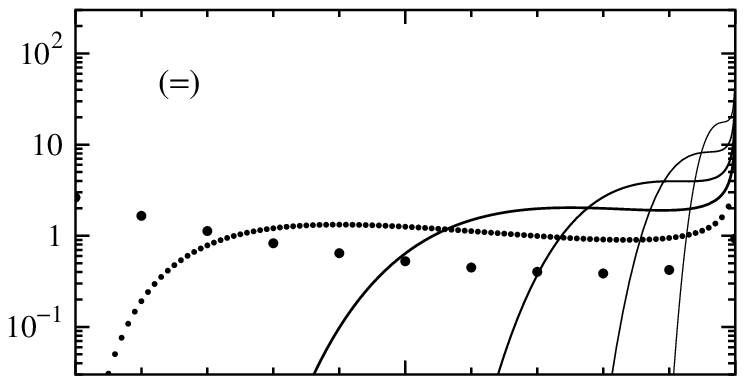}
\\\vspace*{-42pt}
\includegraphics{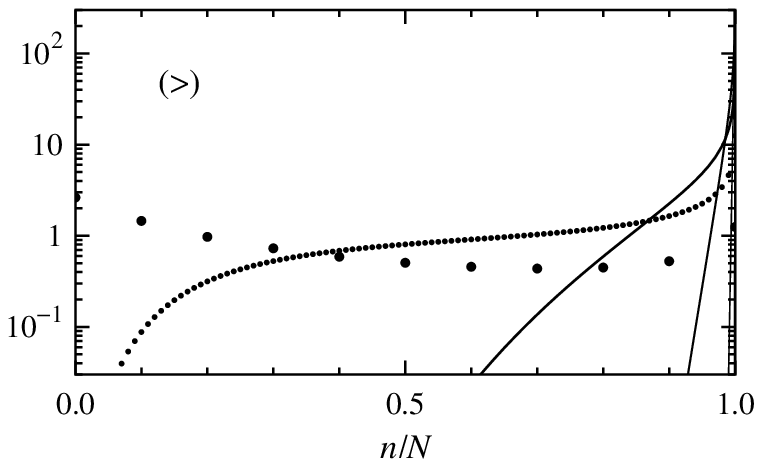}
\end{center}
\caption{\label{fig: K2} The probability density distribution $N\Pn(n)$
  with respect to the fraction of nodes ($n/N$) involved in an
  avalanche. The rule distributions have the same $g(x)$ as displayed
  in Fig.~\ref{fig: gq}, showing rule distributions that are ($<$)
  subcritical, ($=$) critical, and ($>$) supercritical with respect to
  \EP. The displayed networks sizes, $N$, are 10 (large dots), 100
  (small dots), $10^3$ (bold line), $10^4$, $10^5$, and $10^6$
  (gradually thinner lines).}
\end{figure}

\begin{figure}[bt]
\begin{center}
\includegraphics{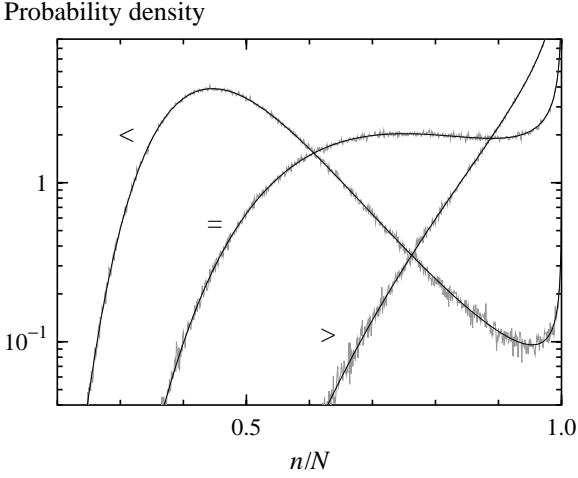}
\end{center}
\caption{\label{fig: K2_num} A numeric comparison between analytic
  calculations (black lines) and explicit reductions of random Boolean
  networks (gray lines). For both cases, the probability density
  distribution $N\Pn(n)$ is displayed as a function of $n/N$. The rule
  distributions have the same $g(x)$ as displayed in Figs.\ \ref{fig:
  gq} and \ref{fig: K2}, showing rule distributions that are ($<$)
  subcritical, ($=$) critical, and ($>$) supercritical with respect to
  \EP. The \UBA\ rule distributions are realized in random Boolean
  networks by rule distributions with the following respective
  selection probabilities: $1/8$, $1/4$, $5/8-p_{\trm r}$, and
  $p_{\trm r}$ for a constant rule, a rule that depends on exactly 1
  input, a canalizing rule that depends on 2 inputs, a 2-input
  reversible rule. The values of $p_{\trm r}$ are ($<$) 0, ($=$) 1/8,
  and ($>$) 1/4. For each rule distribution, $10^6$ networks were
  tested.}
\end{figure}

As can be seen from Fig.~\ref{fig: gq}, a small change in $g(x)$ may
lead to a qualitative change in $G(x)$ for rule distributions close to
criticality.  Such changes have a strong impact on the avalanche size
distribution for large $N$.  Figure~\ref{fig: K2} shows the probability
density distribution of the fraction, $n/N$, of nodes that are
affected by avalanches in networks with the above mentioned rule
distributions. The probability distributions are obtained by recursive
calculation of the distribution of $n_\msa$ as $n_1$ increases. The
recurrence relations are obtained from Eqs.~\eqref{eq: n-update
  first}--\eqref{eq: n-update last} and the result is exact up to
truncation errors. To verify these calculations, we generated
$10^6$ random Boolean networks of size $N=10^3$ for each of the above
described rule distributions. The distributions in the numbers of
frozen nodes in those networks are displayed in Fig.~\ref{fig:
K2_num}.

\begin{figure}[bt]
\begin{center}
\includegraphics{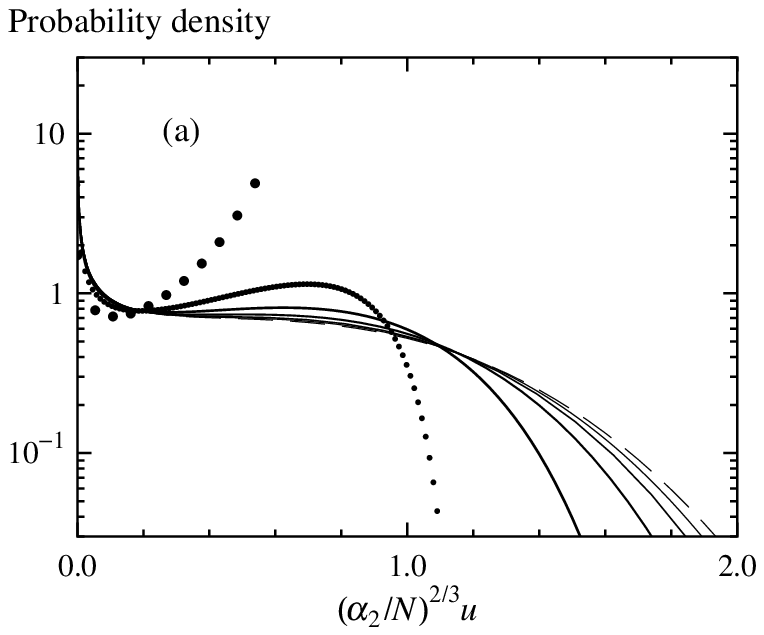}
\includegraphics{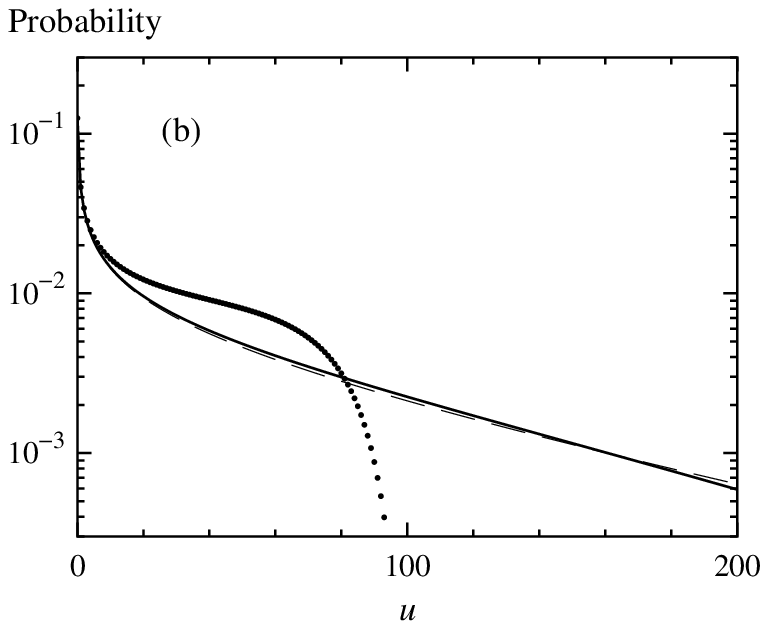}
\end{center}
\caption{\label{fig: EP_K2} Rescaled versions of the probability
  distributions displayed in Fig.~\ref{fig: K2}: (a) the probability
  density for the critical case, with respect to the rescaled number
  of undamaged nodes, $\ut\equiv(\alpha_2/N)^{2/3}u=u/(4N^{2/3})$; (b)
  the probability distribution $\Pu(u)$ for the supercritical case
  without rescaling. The displayed networks sizes, $N$, are 10 (large
  dots), 100 (small dots), $10^3$ (bold line), $10^4$, $10^5$, and
  $10^6$ (gradually thinner lines).  The analytically derived
  asymptotes are shown as dashed lines.  In (b), the distributions for
  networks of sizes $10^4$, $10^5$, and $10^6$ are not plotted because
  they are indistinguishable from the asymptotic curve.}
\end{figure}

In Fig.~\ref{fig: EP_K2}, the probability distributions of the number
of undamaged nodes, $u$, are shown in comparison to the asymptotic
results in Eqs.~\eqref{eq: pu(u) exact asympt} and~\eqref{eq: p(ut) asympt}.
Our analytic results are strengthened by the data in Fig.~\ref{fig:
EP_K2} as the distributions for finite networks approaches the
predicted asymptotes. Finite size effects are clearly visible in the
critical case even for network sizes as big as $N=10^6$, whereas
convergence in the supercritical case is achieved for $N \gtrsim
10^3$.

Kaufman, Mihaljev, and Drossel\ studied distributions of unfrozen nodes in
2-input critical \RBN s using a method similar to ours in that
differential equations for populations of different types of nodes are
developed from a discrete process in which frozen nodes are identified
by the propagation of information from their inputs~\cite{Kaufman:05}.
Their result for the numbers of unfrozen nodes
in 2-input critical \RBN s corresponds to a particular application of
Eq.~\eqref{eq: p(ut) asympt}. In Ref.~\cite{Kaufman:05}, the function
corresponding to $P(\ut)$ [which they call $G(y)$] is
determined by running a stochastic process and a numerically motivated
approximation is proposed:
\begin{align}
  P(\ut) &\approx 0.25\exp(-\tfrac12\ut^3)
     \frac{1-0.5\sqrt{\ut}+3\ut}{\sqrt{\ut}}.
\label{eq: p(ut) Kaufman approx}
\end{align}
The scaling law $P(\ut)\propto\ut^{-1/2}$ for small $\ut$ is also
derived analytically in Ref.~\cite{Kaufman:05}.

For large $x$, Eqs.~\eqref{eq: diffuse}--\eqref{eq: diffuse bound 1}
imply $\pt(0,x)\propto x$ for large positive $x$.  This means that
\begin{align}
  P(\ut) \approx \sqrt{\frac{2\ut}{\pi}}\exp(-\tfrac12\ut^3)
\end{align}
for large $\ut$. Thus the large $\ut$ limit of Eq.~\eqref{eq: p(ut)
  Kaufman approx} differs from the exact result by a factor of
$(3/4)\sqrt{\pi/2}$, an underestimate of about 6\%.

We are able to improve further on Eq.~\eqref{eq: p(ut) Kaufman approx}
by numerical investigations of $\pt(0,x)$ calculated by the
Crank--Nicholson method (see, e.g., \cite{numrecip}) using
Eqs.~\eqref{eq: diffuse}--\eqref{eq: diffuse bound 1}.  We find that
the high-precision numerical results are fit by the function
\begin{align}
  P(\ut) \approx \sqrt{\frac{2\ut}{\pi}}\exp(-\tfrac12\ut^3)
           \biggl(\!1\!+\frac1{3.248\ut+4.27\ut^2+4.76\ut^3}\!\biggr)
\end{align}
with a relative error that is maximally 0.25\% and vanishing for large
$\ut$.

By explicitly keeping track of the populations of nodes with each of
the different types of Boolean logic functions as links from frozen
nodes are deleted, Kaufman, Mihaljev, and Drossel~\cite{Kaufman:05}
also derive results for other quantities, such as the number of links
in the sub-network of unfrozen nodes.  The \EP\ formalism described
above can be applied once again to investigate these additional
quantities in a broader class of networks.  Detailed results for \RBN
s with various degree distributions will be presented elsewhere.

\section{Summary and discussion}

Unordered binary avalanches can in some cases lead to damage on every
node or almost every node of a network, a phenomenon we have dubbed
{\it exhaustive percolation}.  We have studied a broad class of random
networks that can exhibit \EP.  We have shown how to calculate the
probability $\Pex(N)$ that complete coverage occurs (i.e that all
nodes are damaged) and also derived expressions for the probability
distribution $P(u)$ of the number of undamaged nodes, $u$, in the
large $N$ limit when \EP\ does occur.  A logical curiosity in our
approach is the fact that the calculation of $P(u)$ involves
application of the $\Pex$ result to subnetworks containing candidate
sets of damaged nodes.

Our primary results flow from the realization that all of the relevant
information about a \UBA\ defined on a random network is contained in
the damage propagation function $g(x)$ or, equivalently, the damage
control function $q(x)$.  We derive scaling law exponents and exact
results for the distribution of $u$ that are valid for a broad class
of random networks and Boolean rule distributions in the \EP\ regime
and for networks at the \EP\ critical point.  This class includes the
\UBA s that determine the set of frozen nodes in \RBN s with more than
two inputs per node and therefore constitute a generalization of the
results on the set of unfrozen nodes in \RBN s presented in
Ref.~\cite{Kaufman:05}.  Interestingly, the asymptotic behavior found
in Ref.~\cite{Kaufman:05} for the distribution of $u$ at the critical
point is shown to be valid for a broad class of network problems.  

For networks outside the above mentioned class but within the
framework of \UBA, we find connections to previous work on
Galton--Watson processes \cite{Otter:49} and random maps
\cite{Harris:60}.  The central result of our investigations is
displayed in Eqs.~\eqref{eq: pu(u) asympt} and~\eqref{eq: p(ut)
  asympt}, which provide explicit formulas for the probability of
finding $u$ undamaged nodes after an avalanche runs to completion.
The out-degree distributions of the networks described by our formulas
are all Poissonian, but the in-degree distributions may have different
forms, including power laws, so long as the probability of having
in-degree $K$ decays faster than $K^{-3}$.  The exact nature of the
\EP\ transition on networks with broader in-degree distributions is an
interesting issue for future research.  Further work is also needed to
handle correlations between input links to different nodes, a
situation that arises, for example, in random regular graphs or
networks with scale free out-degree distributions.

Our original motivation for studying \EP\ arose from attempts to
understand the dynamical behavior of \RBN s.  We have described one
nontrivial example of how the \EP\ formalism is relevant: the
calculation of the probability distribution for the number of unfrozen
nodes in any \RBN\ with a rule distribution that leads to a given
damage control function $q$ for the associated \UBA.  The problem of
determining how many of the unfrozen nodes are actually relevant for
determining the attractor structure of the \RBN\ can also be framed as
an \EP\ problem, which will be addressed in a separate publication.

\section*{Acknowledgment} This work was supported by the National Science
Foundation through Grant No.~PHY-0417372.

\appendix

\section{Calculation of the damage control function}
\label{app: q calc}

Let $p_K$ denote the probability that a rule has $K$ inputs, and let
$P_0(K,m)$ denote the probability that the output value is zero of a
rule with $K$ inputs fed with $m$ zeros and $K-m$ ones. Then, the
the damage control function is
\begin{align}
  q(x) =\,&\sum_{K=0}^\infty p_K\sum^K_{m=0}P_0(K,m) \binom K m
  x^m(1-x)^{K-m}.
\label{eq: q explicit 0}
\end{align}
Eq.~\eqref{eq: q explicit 0} can be written as
\begin{align}
  q(x) &= a_0 + a_1x + a_2x^2 + \cdots
\label{eq: q explicit 1}
\end{align}
where
\begin{align}
   a_i = \sum_{K=i}^\infty\sum_{m=0}^i p_KP_0(K,m)
                 (-1)^{i-m}\binom K m\binom{K-m}{i-m}.
\label{eq: q expansion terms}
\end{align}
The expansion in Eq.~\eqref{eq: q explicit 1} is well-defined up to
the first term such that the sum in Eq.~\eqref{eq: q expansion terms}
is not absolute convergent. The factor $\binom K m\binom{K-m}{i-m}$
scales like $K^i$ for large $K$ and $P_0(K,m)\le 1$. Hence, $a_i$ is
well-defined if $\sum_{K=0}^\infty K^ip_K$ is convergent and this is
true if $p_K$ decays faster than $K^{-i-1}$.

In addition, the requirement that the output of each rule in the rule
distribution is 1 if all of its inputs have the value 1, yields that
$a_0=0$.  Thus, the expansion
\begin{align}
  q(x) &= \alpha_1x - \alpha_2x^2 + \Oc(x^3),
\label{eq: q-Taylor app}
\end{align}
is valid for all rule distributions such that $p_K$ decays faster than
$K^{-4}$. In the case that $p_K$ decays slower than $K^{-4}$ but
faster than $K^{-3}$, only the residue term can be affected.

\section{Probability for complete coverage}
\label{app: PEP}

Here, we assume that the expansion in Eq.~\eqref{eq: q-Taylor app} is
well-defined. Then, we get
\begin{align}
  \frac{\partial p}{\partial n_\msia}
     &= \frac{c}{2\alpha_1Nn_\msia^2}\frac{\partial^2 p}{\partial c^2}
                        [1+\Oc(n_\msia/N)],
\label{eq: p-diffuse n c}
\intertext{and}
  c_{\trm{max}}(x)/N &= \frac{1}{\alpha_1} 
                         + \frac{\alpha_2}{\alpha_1^2}x 
                         + \Oc(x^2).
\end{align}

To remove the dependence of $n_\msia$ from the leading order term of
the diffusion rate in Eq.~\eqref{eq: p-diffuse n c}, we let
$t=-1/n_\msia$. By also letting $y=1-\alpha_1c/N$, we rewrite
Eq.~\eqref{eq: p-diffuse n c} to a form that easily can be rescaled
as $N$ grows. We get
\begin{align}
  \frac{\partial p}{\partial t} &= \frac{1-y}{2}
         \frac{\partial^2 p}{\partial y^2}[1+{\cal O}(\tfrac1{Nt})].
\intertext{and}
  y_{\trm{min}} &= -\frac{\alpha_2}{\alpha_1Nt}[1+\Oc(\tfrac1{Nt})],
\label{eq: ymin}
\end{align}
where $y=y_{\trm{min}}$ is the transformed value of $c_{\trm{max}}$.
The boundary conditions are $p=0$ for $y=y_{\trm{min}}$ and $p=1$
for $y=1$.

The $N$ dependence of the leading order term of the boundary condition
in Eq.~\eqref{eq: ymin} can be removed by rescaling of $y$ and $t$.
Typically, $\alpha_2>0$ and this is the case that we will focus on.
[Note that Eq.~\eqref{eq: x0 gt} means that $\alpha_2$ must be
nonnegative at the transition.] If $\alpha_2=0$, either $q(x)=x$ or
$q(x)=x-\alpha_m x^m+\cdots$ with $m>2$ (apart from some pathological
special cases). The first case, $q(x)=x$, is a special case that is
convenient for analytic calculation, whereas the latter case require
calculations analogous to the calculations for $\alpha_2>0$. We will
come back to the case $q(x)=x$.

For $\alpha_2>0$, we rescale $y$ and $t$ according to
\begin{align}
  \yt &= \Nt^{1/3}y
\intertext{and}
  \tti&= \Nt^{2/3}t,
\intertext{where}
  \Nt &= \frac{\alpha_1}{\alpha_2}N.
\label{eq: Nt}
\end{align}
Then,
\begin{align}
  \frac{\partial p}{\partial\tti} &= \frac12
         \frac{\partial^2p}{\partial\yt^2}
                \bigl(1+\Nt^{-1/3}\yt\bigr)
                \bigl[1+{\cal O}\bigl(\tfrac1{\Nt^{1/3}\tti}\bigr)\bigr],
\label{eq: p diff tt yt}
\intertext{where}
  \yt_{\trm{min}} &= \tti^{-1}
                \bigl[1+{\cal O}\bigl(\tfrac1{\Nt^{1/3}\tti}\bigr)\bigr].
\end{align}

The boundary conditions are $p=0$ for $\yt_{\trm{min}}$ and $p=1$ for
$y=\Nt^{1/3}$. The only plausible limit of $p$ as
$t\rightarrow-\infty$ is $p=y/\Nt^{1/3}$. To get a motivation that is
mathematically acceptable, one needs to relate the original integer
based formulation of the problem in Eqs.\ \eqref{eq: n-update
first}--\eqref{eq: n-update last}. The large $N$ behavior of
Eq.~\eqref{eq: n-update last},
\begin{align}
  \lim_{N\rightarrow\infty}U(\s,j) &= 1/n_\msia,
\label{eq: a lim}
\end{align}
yields
\begin{align}
  \lim_{N\rightarrow\infty}\Pex(N,q;n_0,n_\msa)
             &= \frac{n_\msa}{n_\msia}.
\label{eq: Pex n fixed N large}
\end{align}
Eq.~\eqref{eq: Pex n fixed N large} can be shown via induction. The
induction is initiated by
\begin{align}
  \lim_{N\rightarrow\infty}\Pex(N,q;1,0)&=0
\intertext{and}
  \lim_{N\rightarrow\infty}\Pex(N,q;0,1)&=1,
\end{align}
which means that Eq.~\eqref{eq: Pex n fixed N large} is true for
$n_\msia=1$. To obtain the induction step, we assume that
Eq.~\eqref{eq: Pex n fixed N large} is true for
$n'_\msia=n_\msia-1$. Then, we get
\begin{align}
  \lim_{N\rightarrow\infty}\Pex(N,q;n'_0,n'_\msa)
             &= \frac{n'_\msa}{n_\msia-1}
\end{align}
which leads to
\begin{align}
  \lim_{N\rightarrow\infty}\Pex(N,q;n_0,n_\msa) 
             &= \frac{\langle n'_\msa\rangle}{n_\msia-1}\\
             &= \frac{n_\msa + n_0/n_\msia-1}{n_\msia-1}\\
             &= \frac{n_\msa}{n_\msia}
\end{align}
that completes the induction step. Eq.~\eqref{eq: Pex n fixed N large}
means that the value of $p$ approaches a linear function of $\yt$ for
$\tti=N^{2/3}/n_0$ as $N\rightarrow\infty$. Hence, the boundary
condition for $t\rightarrow-\infty$ is $p=y/\Nt^{1/3}$.

Rescaling of $p$ according to
\begin{align}
  \pt &= \Nt^{1/3}p
\end{align}
gives the boundary condition
\begin{align}
  \lim_{t\rightarrow-\infty}\pt = \yt.
\label{eq: p t neg inf}
\end{align}
If Eq.~\eqref{eq: p t neg inf} is extended to be valid for all
non-negative $\yt$, the boundary condition at $\yt=\Nt^{1/3}$ can be
dropped. In the limit of large $N$, Eq.~\eqref{eq: p diff tt yt}
becomes
\begin{align}
  \frac{\partial\pt}{\partial\tti} &= \frac12
         \frac{\partial^2\pt}{\partial\yt^2}.
\label{eq: app diffuse}
\end{align}
With $\pt$ is written on the form $\pt(\tti,\yt)$, the boundary
conditions are
\begin{align}
  \pt(\tti,1/\tti) &= 0 & \trm{for }&\tti<0
\label{eq: app diffuse bound 0}
\intertext{and}
  \lim_{\tti\rightarrow-\infty}\pt(\tti,\yt) &= \yt
           & \trm{for }&\yt\ge0,
\label{eq: app diffuse bound 1}
\end{align}
as $N\rightarrow\infty$.

The solution to Eqs.~\eqref{eq: app diffuse}--\eqref{eq: app diffuse bound 1}
can be calculated numerically. By expressing the transformed variables
$\tti$ and $\yt$ in terms of more fundamental quantities, we get
\begin{align}
  \Pex&(N,q; n_0,n_\msa) \nonumber\\  
        &\approx \Nt^{-1/3}\pt\biggl[-\frac{\Nt^{2/3}}{n_\msia},
                  \Nt^{1/3}\biggl(1-\frac{\alpha_1n_0}
                                        {Nq(n_\msia/N)}\biggr)\biggr]
\label{eq: Ploopfree rescaled}
\end{align}
and
\begin{align}
  \Pex&(N,q; N-n_\msa,n_\msa)  \nonumber\\  
        &\approx \Nt^{-1/3}\pt\biggl[-\frac{\Nt^{2/3}}{N},
                  \Nt^{1/3}\biggl(1-\frac{\alpha_1n_0}{Nq(1)}\biggr)\biggr]
\end{align}
for large $N$.

If the avalanche is initiated by letting each node start from 0 with
probability $q(1)$, we get
\begin{align}
  \langle n_\msa^\ti\rangle &= N[1-q(1)]
\intertext{and}
  \sigma(n_\msa^\ti) &= \sqrt{Nq(1)[1-q(1)]}.
\end{align}

Provided that $q(x)$ does not depend on $N$, $\Nt$ is fixed and the
spread in $\yt$ that correspond to the initial value of $m$ will go to
zero as $N\rightarrow\infty$.  Also, $\Nt^{2/3}/N$ approaches zero as
$N\rightarrow\infty$.  In this case, the probability for an avalanche
to yield complete coverage is given by
\begin{align}
  \Pex(N, q) &\approx \Nt^{-1/3}\pt[0,\Nt^{1/3}(1-\alpha_1)],
\label{eq: app PexNq}
\end{align}
for large $N$. Only the first two arguments to $\Pex$ are kept in
Eq.~\eqref{eq: app PexNq}, because the process is fully determined by $N$
and $q$.

In the special case that $q(x)=x$ for all $x\in[0,1]$, Eq.~\eqref{eq:
n-update last} yields $a=1/n_\msia$, which is a strong form of
Eq.~\eqref{eq: a lim}. By using the same induction steps that lead
from Eq.~\eqref{eq: a lim} to Eq.~\eqref{eq: Pex n fixed N large}, we
conclude that
\begin{align}
   \Pex(N,x\mapsto x;n_0,n_\msa)
             &= \frac{n_\msa}{n_\msia}.
\label{eq: q-ident}
\end{align}

\section{Asymptotes for sparse percolation}
\label{app: SP}

Provided that the derivative of $g(x)$ is well defined at $x=0$, we
let $\lambda = g'(0)$, where $g'(x)$ denotes the derivative of
$g(x)$. Then,
\begin{align}
  \lim_{N\rightarrow\infty} g^1_{N,n}(x) &= x,
\intertext{which means that}
  \lim_{N\rightarrow\infty} q^1_{N,N-n}(x) &= x
\intertext{and}
  \lim_{N\rightarrow\infty} \PNex(n,\ell;N) &= \frac\ell n
\end{align}
according to Eq.~\eqref{eq: q-ident}. Thus, the large $N$ limit of
Eq.~\eqref{eq: Pn1} is
\begin{align}
  \lim_{N\rightarrow\infty}\Pn(n)
      &= \rho(n)\tau(n,\ell)\lambda^{n-\ell}e^{n(1-\lambda)}\frac\ell n\\
      &= \frac{\ell(n\lambda)^{n-\ell}}{n(n-\ell)!}e^{-n\lambda}.
\label{eq: lim Pn}
\end{align}
For large $N$, Eq.~\eqref{eq: lim Pn} yields
\begin{align}
  \lim_{N\rightarrow\infty}\Pn(n)
     &\approx \frac \ell{\sqrt{2\pi}}e^{n(1-\lambda)}\lambda^{n-\ell}n^{-3/2}.
\end{align}

Due to the correspondence to well investigated branching processes,
Eq.~\eqref{eq: lim Pn} is not a new result. For the special case of
$\ell=1$, Eq.~\eqref{eq: lim Pn} is given explicitly in
Ref.~\cite{Otter:49}, and the general form of Eq.~\eqref{eq: lim Pn}
can easily be obtained by the theorem presented in
Ref.~\cite{Dwass:69}.

\section{Asymptotes for exhaustive percolation}
\label{app: EP asympt}

In analogy with our investigation of \SP, we assume that $q(x)$ has a
well-defined derivative at $x=0$ and let $\lambda = q'(0)$. For \EP\ to
be likely in the large $N$ limit, it is required that $q(x)\leq x$ for
all $x$, meaning that $\lambda\le1$. The large $N$ behavior of
Eq.~\eqref{eq: Pu0} is partly explained by
\begin{align}
  \lim_{N\rightarrow\infty} \frac{\rho(u)\rho(N-u)}{\rho(N)}[Q(u/N)]^N
     &= \rho(u)\lambda^ue^{u(1-\lambda)},
\label{eq: lim EP}
\end{align}
but it remains to investigate the role of $\PNex(N-u,0;N)$. To this end,
we consider the ratio $\PNex(N-u,0;N)/\Pex(N)$. [Here, we have dropped the
argument $q$ from $\Pex(N, q)$.]

When $N\rightarrow\infty$, there are two processes that influence on
this ratio: $q^1_{N,u}$ approaches $q$ and $N$ increases. The increase
of $N$ makes the involved probabilities more sensitive for the
shrinking differences between $q^1_{N,u}$ and $q$. Thus, there are two
competing processes as $N\rightarrow\infty$.  The sensitivity with
respect to $q$ is limited by the variance of the number of nodes with
initial state \tsa, because this variance can be seen as a rescaling
of $q$. The change in $q(x)$ by such a rescaling scales like
$q(x)/\sqrt{N}$ for large $N$. If $q(x)$ has a well-defined nonzero
derivative at $x=0$, the difference $q^1_{N,u}(x)-q(x)$ scales like
$q(x)/N$ for large $N$. Hence, the decrease in the difference between
$q^1_{N,u}$ and $q$ dominates over the increase in sensitivity,
meaning that
\begin{align}
  \lim_{N\rightarrow\infty} \frac{\PNex(N-u,0;N)}{\Pex(N)}
     &= 1.
\end{align}
Thus,
\begin{align}
  \lim_{N\rightarrow\infty}\frac{\Pu(u)}{\Pex(N)}
     &= \rho(u)\lambda^ue^{u(1-\lambda)}\\
     &= \frac{(u\lambda)^{u}}{u!}e^{-u\lambda},
\label{eq: Pu1}
\end{align}
where $(u\lambda)^u$ should be interpreted with the convention that
$0^0=1$ in order to handle the case $u = 0$ properly.

\subsection{Limit distributions for supercritical EP}
\label{app: super EP}

If $0<\lambda<1$ and $x=0$ is the only solution to $q(x)=x$ in the
interval $0\le x\le1$, the exponential decay of $[Q(u/N)]^N$, in
Eq.~\eqref{eq: Pu0}, with increasing $u$ ensures that
\begin{align}
  \sum_{u=0}^\infty\lim_{N\rightarrow\infty}\Pu(u) &= 1
\intertext{and}
  \Bigl[\lim_{N\rightarrow\infty}\Pex(N)\Bigr]\sum_{u=0}^\infty
       \frac{(u\lambda)^{u}}{u!}e^{-u\lambda} &= 1.
\end{align}
Thus, $\lim_{N\rightarrow\infty}$ has a unique value for each
$\lambda$. This value can be calculated by considering the simplest
case, $q(x)=\lambda x$.  From the definition of the spreading
process, we get
\begin{align}
  \Pex(N,x\mapsto\lambda x;n_0,n_\msa) &= \Pex(N,x\mapsto x;n_0,n_\msa).
\end{align}
Then, Eq.~\eqref{eq: q-ident} and averaging over initial configurations
yield
\begin{align}
  \Pex(N,x\mapsto\lambda x) &= 1-\lambda,
\end{align}
which means that
\begin{align}
  \lim_{N\rightarrow\infty}\Pex(N, q) &= 1-\lambda~
\end{align}
for all $q$ that satisfy the above mentioned criteria. We get,
\begin{align}
  \lim_{N\rightarrow\infty}\Pu(u) 
        &= (1-\lambda)\frac{(u\lambda)^{u}}{u!}e^{-u\lambda},
\label{eq: P(u) asympt}
\intertext{which for large $u$ means that}
  \lim_{N\rightarrow\infty}\Pu(u) 
        &\approx\frac{1-\lambda}{\sqrt{2\pi}}e^{u(1-\lambda)}
            \lambda^uu^{-1/2}.
\end{align}

\subsection{Scaling at the EP transition}
\label{app: crit EP}

This section aims to derive the asymptotic distribution of $u$ for
large $N$ for critical \EP\ with $\alpha_2>0$ and $\ell=0$. Define
$\alpha_1^1$, $\alpha_2^1$ and $\Nt^1$ analogous to the definitions of
$\alpha_1$, $\alpha_2$ and $\Nt$ in Eqs.~\eqref{eq: q-Taylor} and
\eqref{eq: Nt}. The derivatives of $q^1(x)$ in Eq.~\eqref{eq: q1} at
$x=0$ are given by
\begin{align}
  (q^1)'(0) &= q'(u/N)\frac{1-u/N}{1-q(u/N)}
\intertext{and}
  (q^1)''(0) &= q''(u/N)\frac{(1-u/N)^2}{1-q(u/N)}
\end{align}
and we get
\begin{align}
  \Nt^1 &=\Nt+\Oc(u/N)
\intertext{and}
  \alpha_1^1 &= \alpha_1 + (\alpha_1-1-2\alpha_2)u/N
                       +\Oc(u^2/N^2).
\end{align}
For critical networks, with $\alpha_1=1$, we get $\Nt = N/\alpha_2$ and
\begin{align}
  \alpha_1^1 &= 1 - 2u/\Nt + \Oc(u^2/N^2).
\end{align}
Insertion into Eq.~\eqref{eq: app PexNq} yields
\begin{align}
  \PNex(N-u,0;N)
    &\approx \Nt^{-1/3}\pt(0,2\Nt^{-2/3}u)
\end{align}
for $u\ll N$.

Because $Q(x)\leq1$ with equality if and only if $q(x)=x$, $\Pu$
that vanishes exponentially, as $N$ goes to infinity, for any fixed
$u/N$ such that $q(u/N)\ne u/N$. For a typical network with
$\alpha_2>0$ at the \EP\ transition, the only solution to $q(x)=x$ is
$x=0$. For such a network, the large $N$ behavior of $\Pu$ is found
by expanding $Q(x)$ around $x=0$. To the leading
non-vanishing order, we get
\begin{align}
  Q(x) &\approx \exp\biggl(-\frac{\alpha_2^2x^3}2\biggr),
\end{align}
which yields
\begin{align}
  \Pu(u) 
     &\approx \rho(u)\exp\biggl(-\frac{u^3}{2\Nt^2}\biggr)
       \PNex.
\end{align}

Hence,
\begin{align}
  \Pu(u) &\approx \Nt^{-1/3}\rho(u)
                \exp\biggl(-\frac{u^3}{2\Nt^2}\biggr)
                \pt(0,2\Nt^{-2/3}u),
\end{align}
with asymptotic equality for large $N$. The probability density,
$P(\ut)$, for the distribution of $\ut=\Nt^{-2/3}u$ as
$N\rightarrow\infty$ approaches
\begin{align}
  P(\ut) &= \frac{\exp(-\frac12\ut^3)}{\sqrt{2\pi\ut}}
                        \,\pt(0,2\ut).
\end{align}\\

\section{Exact results}
\label{app: exact}

A network with $g(x)=x$ is critical for all $x$. For such a network,
$G(x)=1$ for all $x$ and $\PNex(n,\ell;N)=\ell/n$. Hence, Eq.~\eqref{eq: Pn1}
yields
\begin{align}
  \Pn(n) 
      &= \frac{\rho(n)\rho(N-n)}{\rho(N)}\frac{\tau(n, \ell)}
               {\tau(N, \ell)}\frac\ell n\\
      &= \frac\ell n\binom{N-\ell}{n-\ell}\frac{n^{n-\ell}
           (N-n)^{N-n}}{N^{N-\ell}}.
\label{eq: Pn ident}
\end{align}
For $n$ and $N$ satisfying $n\gg1$ and $N-n\gg1$, we get
\begin{align}
  \Pn(n) &\approx \frac \ell{\sqrt{2\pi}}\frac{\sqrt{N}}{n\sqrt{n(N-n)}}\,.
\end{align}

In the special case $\ell=1$, $\Pn(n)$ is the distribution of the number
of predecessors to an element in a random map. This distribution,
which is consistent with eq.~\eqref{eq: Pn ident} for $\ell=1$, was
obtained in Ref.\ \cite{Rubin:54} and restated in Ref.\
\cite{Harris:60}.

For completeness, we provide an explicit expression for the
distribution of avalanche sizes in the case that $g(x)$ is a first
order polynomial. From Eq.~\eqref{eq: Pn1}, we get
\begin{align}
  \Pn(n)
      =\,& \Pu(u) \nonumber\\  
      =\,&
                     \frac{n^{n-\ell}u^u}{N^{N-\ell}}\binom{N-\ell}{n-\ell}
                     \biggl(g(0)+\frac\ell n\biggr)
      \nonumber\\&\times
                     \biggl(\frac{ug(0)+ng(1)}n\biggr)^{\!n-\ell}
                     \biggl(\frac{nq(0)+uq(1)}u\biggr)^{\!u}.
\label{eq: explicit Pn}              
\end{align}

\end{document}